\documentclass[aps,prx,twocolumn,longbibliography,groupedaddress,floatfix,superscriptaddress]{revtex4-1}
\usepackage{graphicx}  
\usepackage{dcolumn}   
\usepackage[english]{babel}
\usepackage{bm}        
\usepackage{amsmath, amsthm, amssymb}
\usepackage{bbold} 
\usepackage{mathrsfs}
\usepackage{array}
\usepackage[usenames]{color}
\usepackage{xr}
\usepackage{soul} 
\usepackage{ulem}
\usepackage{comment}
\definecolor{amber}{rgb}{1.0, 0.49, 0.0}

\emergencystretch=\maxdimen
\begin{document}

\title{Charge Density Wave and Superconductivity in the
Disordered Holstein Model}
\author{B. Xiao} 
\affiliation{Department of Physics, University of California,
Davis, California 95616, USA}
\affiliation{Center for Computational Quantum Physics, Flatiron Institute, 
New York, New York 10010, USA} 
\author{N.C. Costa} 
\affiliation{Instituto de F\'\i sica, Universidade Federal do Rio de Janeiro
Cx.P. 68.528, 21941-972 Rio de Janeiro RJ, Brazil}
\affiliation{International School for Advanced Studies (SISSA),
Via Bonomea 265, 34136, Trieste, Italy}
\author{E. Khatami} 
\affiliation{Department of Physics and Astronomy, San Jos\'{e} State
University, San Jos\'{e}, California 95192, USA}
\author{G. G. Batrouni}
\affiliation{Universit\'e C\^ote d'Azur, CNRS, INPHYNI, 0600 Nice, France}
\affiliation{Centre for Quantum Technologies, National University of
  Singapore, 2 Science Drive 3, 117542 Singapore} 
\affiliation{Department of Physics, National University of Singapore, 2
  Science Drive 3, 117542 Singapore}
\affiliation{Beijing Computational Science Research Center, Beijing
  100193, China}
\author{R.T. Scalettar}
\affiliation{Department of Physics, University of California, 
Davis, California 95616, USA}

\begin{abstract}
The interplay between electron-electron correlations and disorder has
been a central theme of condensed matter physics over the last several
decades, with particular interest in the possibility that interactions
might cause delocalization of an Anderson insulator into a metallic
state, and the disrupting effects of randomness on magnetic order and
the Mott phase. Here we extend this physics to explore {\it
  electron-phonon} interactions and show, via exact quantum Monte
Carlo simulations, that the suppression of the charge density wave
correlations in the half-filled Holstein model by disorder can
stabilize a superconducting phase.  
Our simulations thus capture qualitatively the 
suppression of charge ordered phases
and emergent superconductivity 
recently seen experimentally.
\end{abstract}


\maketitle

\noindent
{\it Introduction.}
Although the problem of the localizing effect of randomness
on {\it non-interacting} electrons is well understood
\cite{abrahams79,wegner80,efetov80},
the combined effects of disorder and electron-electron
interactions 
remain an area of continued theoretical and 
experimental interest
\cite{lee85,giamarchi88,belitz94,dagotto05,balatsky06,abrahams10,
dobrosavljevic12,Vojta19}.
A traditional focus has been on the possibility of electron-electron
interactions inducing an insulator-to-metal transition in two dimensions
\cite{Kravchenko03},
but recent
attention has also turned to understanding the interplay
in the context of modern developments including
Majorana fermions~\cite{lobos12},
topological bands~\cite{krishna19},
ultracold atomic gases~\cite{kondov11},
and many-body localization
\cite{basko06,nandkishore15,barlev15}.
Supplementing analytic calculations,
numerical approaches have attempted to address the issue with
techniques which treat disorder and electronic correlations
non-perturbatively
\cite{Denteneer99,terletska18}.
Unfortunately, in quantum Monte Carlo (QMC) methodologies,
the combination of 
randomness and interactions often leads to the fermion minus-sign problem, 
a bottleneck which dramatically limits their
effectiveness~\cite{loh90,troyer05,iglovikov15}.

In this work, we use an exact sign-problem-free QMC approach to investigate the
interplay between randomness and {\it electron-phonon interactions}.  
This is an area far less explored with numerical simulations
than that of randomness and electron-electron interactions.  
This gives us the opportunity, within the framework of the disordered Holstein model, 
to address important fundamental qualitative issues. Among them, we find 
the emergence of a superconducting (SC) phase upon the suppression of
the charge-density wave (CDW) order by randomness.  
Further, the absence of the sign problem allows us to reach
low temperatures, and thus use the full power of QMC calculations
which cannot be fully exploited for electron-electron interactions.

This paper is organized as follows: After describing our 
Hamiltonian and methodology in the ``Model'' and 
``Methods'' sections, respectively, we show in the ``Results'' section 
the details of the quantum simulations which lead to a 
demonstration of the emergence of a SC phase driven by
the interplay of electron-phonon interaction and randomness.
Our final remarks are in the ``Concluding remarks'' section.
Further results about the magnitude of SC and CDW correlations in 
the full temperature-disorder plane are presented in the Supplemental Materials.

\color{black}
\vskip0.02in \noindent
{\it Model.}
The Holstein model describes itinerant electrons whose site density
couples to the displacement of a local phonon mode.  Its Hamiltonian
reads
\begin{align} \label{eq:Holst_hamil}
\nonumber \mathcal{H} = & -t \sum_{\langle \mathbf{i}, \mathbf{j}
  \rangle, \sigma} \big(d^{\dagger}_{\mathbf{i} \sigma}
d^{\phantom{\dagger}}_{\mathbf{j} \sigma} + {\rm h.c.} \big) -
\sum_{\mathbf{i}, \sigma} (\mu^{\phantom{\dagger}} -
\epsilon^{\phantom{\dagger}}_{\mathbf{i}} )
n^{\phantom{\dagger}}_{\mathbf{i}, \sigma} \\ & + \omega_{0} \sum_{
  \mathbf{i} } a^{\dagger}_{\mathbf{i}}
a^{\phantom{\dagger}}_{\mathbf{i}} + g \sum_{\mathbf{i}, \sigma}
n^{\phantom{\dagger}}_{\mathbf{i}\sigma} \big(
a^{\dagger}_{\mathbf{i}} + a^{\phantom{\dagger}}_{\mathbf{i}} \big) ~~
,
\end{align}
in which the sum over $\mathbf{i}$ is on a two-dimensional square
lattice, with $\langle \mathbf{i}, \mathbf{j} \rangle$ denoting
nearest-neighbors.  $d^{\dagger}_{\mathbf{i} \sigma}$
($d^{\phantom{\dagger}}_{\mathbf{i} \sigma}$) is the creation
(annihilation) operator of electrons with spin $\sigma$ at site
$\mathbf{i}$, with $n^{\phantom{\dagger}}_{\mathbf{i}\sigma}\equiv
d^{\dagger}_{\mathbf{i} \sigma} d^{\phantom{\dagger}}_{\mathbf{i}
  \sigma}$ denoting the number operator.
$a^{\dagger}_{\mathbf{i}}(a^{\phantom{\dagger}}_{\mathbf{i}})$ is the
phonon creation (annihilation) operator.  The first term on the right
hand side of Eq.~\eqref{eq:Holst_hamil} corresponds to the hopping of
electrons, and the second term contains the global chemical potential
$\mu$.  Disorder effects are introduced in the second term, by means
of random on-site energies $\epsilon_{\mathbf{i}}$, chosen uniformly
in the range $[-\Delta/2, \Delta/2]$, so that $\Delta/t$ represents
the dimensionless disorder strength.  Local phonon modes, with energy
$\omega_{0}$, are included in the third term.  Finally, the last term
describes their coupling to electrons, with strength $g$.

It is worth noticing that the square lattice dispersion relation has 
a number of special features, such as a perfect nesting and a van-Hove singularity 
in the density of states (at half-filling), which lead to CDW order at weak 
electron-phonon coupling.  For stronger coupling cases, the occurrence of CDW 
order is less dependent on the Fermi surface features, and its behavior 
on a square lattice is generic, e.g.~with CDW transition
temperatures being similar to those on other 
2D bipartite lattices~\cite{Weber18,Zhang19,Chen19,CohenStead20}.
In this work, we analyze both weak and strong coupling regimes
at half-filling, $\langle n_{\mathbf{i \sigma}}\rangle =1/2$, 
which is obtained by fixing $\mu=-2g^{2}/\omega_{0}$,
regardless of the lattice size or temperature, due to an appropriate
particle-hole symmetry.  We further set $t=1$ to represent the
unit of energy, and use units where $\hbar=k_B=1$.  We also define
$\lambda_{D}= g^2/ (z t {\omega_{0}})$ as the dimensionless
electron-phonon coupling, where $z=4$ is the coordination number for
the square lattice.
In what follows, we consider two cases: [i] the
adiabatic case, with $\omega_{0}/t=1/2$ and an intermediate coupling
strength $\lambda_D=1/2$ ($g=1$); and [ii] the anti-adiabatic case,
with $\omega_{0}/t =4$ and a weak coupling strength $\lambda_D=1/4$
($g=2$).

\vskip0.02in \noindent
{\it Methods.} 
We employ the determinant quantum Monte Carlo (DQMC)
method~\cite{Blankenbecler81,Scalettar89,Noack91,Santos03}, an
unbiased auxiliary-field approach that provides finite-temperature
properties of interacting fermions.  Within this approach, both
equal-time and unequal-time quantities can be calculated.  See ~\cite{supplemental} for more details.  

Charge modulations are probed by
analyzing the density-density correlation functions $\langle
n_{\mathbf{i}} n_{\mathbf{j}} \rangle$, and their Fourier transform,
the charge structure factor
\begin{align}
S(\mathbf{q})= \frac{1}{N} 
\sum_{\mathbf{i},\mathbf{j}} 
e^{i \mathbf{q} \cdot \mathbf{(r_{i}-r_{j})}} \langle n_{\mathbf{i}}
n_{\mathbf{j}} \rangle,
\label{eq:Sq}
\end{align}
where $N=L^2$ is the number of lattice sites in the system.
Similarly, superconducting properties are examined by means of the
$s$-wave pairing susceptibility,
\begin{align}
\chi_{s} = \frac{1}{N} \int^{\beta}_{0} \mathrm{d}\tau ~
\langle \Delta(\tau) \Delta^{\dagger}(0) \rangle,
\label{eq:chipairing}
\end{align}
in which $\beta=1/T$ is the inverse temperature and $\Delta(\tau) =
\sum_{\mathbf{i}} d^{\phantom{\dagger}}_{\mathbf{i}\downarrow}(\tau)
d^{\phantom{\dagger}}_{\mathbf{i}\uparrow}(\tau)$, with
$d^{\phantom{\dagger}}_{\mathbf{i}\sigma}(\tau) = e^{\tau \mathcal{H}}
d^{\phantom{\dagger}}_{\mathbf{i}\sigma} e^{-\tau \mathcal{H}}$.
Although the equal-time pairing correlations
at large spatial separation can also be used to probe superconductivity,
the full susceptibility provides a more sensitive measure, especially in the case of 
a Kosterlitz-Thouless transition, as expected to occur in 2D lattices~\cite{Noack91,Huscroft97,Paiva04}.

Finally, we investigate transport properties by calculating a proxy of
the direct current (dc) conductivity~\cite{Trivedi96,Denteneer99}
\begin{equation}
\sigma_{dc} \approx \frac{\beta^2}{\pi} \Lambda_{xx}(\mathbf{q=0}, \tau = \beta/2),
\label{eq:sigma_dc}
\end{equation}
where $\Lambda_{xx}(\mathbf{q}, \tau ) = \langle
j_{x}(\mathbf{q},\tau) j_{x}(-\mathbf{q}, 0) \rangle$ is the
current-current correlation function, and $ j_{x}(\mathbf{q}, \tau) $
is the Fourier transform of 
$j_{x}({\bf r},\tau) = -i \,t\,
\big( d^{\dagger}_{{\bf r+\hat{x}},\sigma} d_{{\bf r},\sigma}^{\phantom{\dagger}}
- d^{\dagger}_{{\bf r},\sigma}
d_{{\bf r+\hat{x}},\sigma}^{\phantom{\dagger}} \big)(\tau)$.
We carry out the calculations on lattices sizes 
from $6\times6$ to $12\times12$, and average our expectation 
values over 110 disorder realizations.

\begin{figure}[t]
\hspace*{-0.6cm}
\includegraphics[scale=0.32]{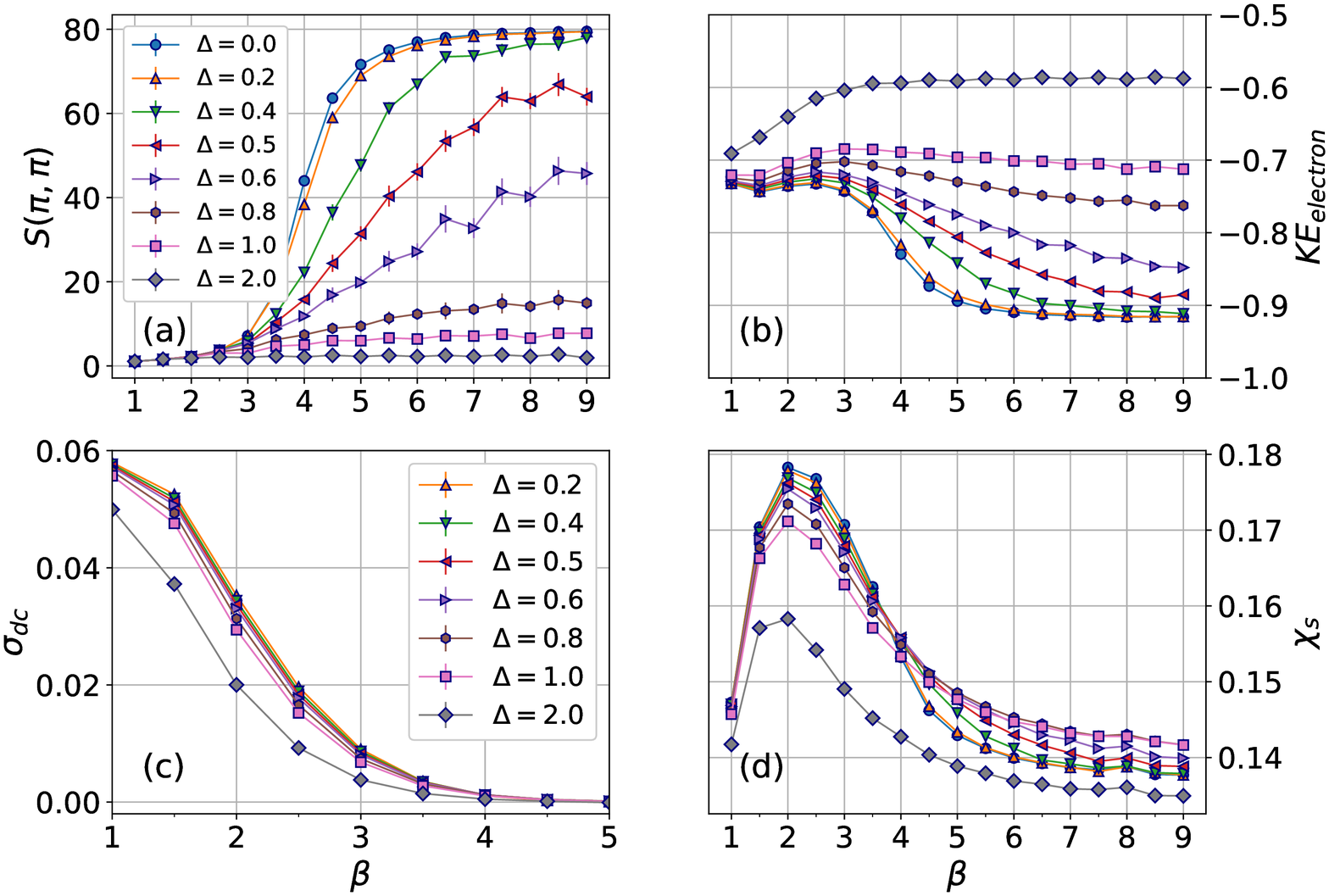}
\caption{The (a) charge structure factor, (b) kinetic energy of
  electrons (c) dc conductivity and (d) s-wave pair susceptibility as
  functions of the inverse temperature, and for different disorder
  strength, at fixed $L=10$, $\omega_0=0.5$, and $\lambda_{D}=0.5$
  ($g=1$).  Results are shown for the dc conductivity only for larger
  $\Delta$, where Eq.~\ref{eq:sigma_dc} is valid\cite{Trivedi96}.
  }
\label{fig:adiabat} 
\end{figure}

\vskip0.02in \noindent
{\it Results.}
We first consider the response of charge modulations to disorder in
the adiabatic case, by fixing $\omega_{0}/t=0.5$ and $\lambda_D=1/2$
($g=1$).  When $\Delta=0$, there is a large enhancement of
$S(\pi,\pi)$ around $\beta\approx 4$, as presented in
Fig.\,\ref{fig:adiabat}\,(a), in line with recent
studies~\cite{Chen18,Shaozhi19} that show a CDW transition at
$\beta_{c} = 4.1 \pm 0.1$ (see also SM).  In presence of weak
disorder, $\Delta \lesssim 0.3t$, the behavior of $S(\pi,\pi)$ is only
slightly changed from that of the clean system, suggesting the
continued existence of long-range charge correlations over length
scales up to the lattice sizes being simulated, as displayed in
Fig.\,\ref{fig:adiabat}\,(a).  However, as disorder increases further,
$S(\pi,\pi)$ has its characteristic energy scale shifted to larger
$\beta$ (lower temperature), and its strength reduced. Eventually, for
$\Delta \approx t$, long-range correlations seem entirely destroyed,
even at very low temperatures.

At this point, it is convenient to estimate the size of $\Delta$ needed 
to break charge order.  From a second order perturbation theory~\cite{Berger95}, 
the effective attraction between electrons is given by $U_{\rm eff} =-2g^2/\omega_0$, 
therefore the CDW scale may be estimated as $4t^2/|U_{\rm eff}| = 2 t^2 \omega_0/g^2$.
Given this, when $\Delta$ exceeds some fraction of this value, one should expect 
the charge correlations to be suppressed.
Indeed, this yields $\Delta_c \lesssim 1$ for $\omega_0 = 0.5, \, g=1$,
in rough agreement with the vanishing of the CDW correlations 
for $\Delta \gtrsim 0.5$, displayed in Fig.\,\ref{fig:adiabat}\,(a).

Further insight into this crossover is provided by the behavior of 
the electronic kinetic energy, exhibited in Fig.\,\ref{fig:adiabat}\,(b).
At weak disorder, despite the occurrence of a Peierls-like
charge gap, the alternation of empty and doubly occupied sites
associated with strong CDW correlations promotes charge fluctuations,
and hence the magnitude of the kinetic energy increases as the
temperature is lowered.  By contrast, in the strong disordered case,
the pairs are localized randomly, with some doublons at adjacent
sites, precluding virtual hopping. As a consequence, the kinetic
energy decreases in magnitude as $T\rightarrow 0$.
Despite the suppression of the CDW order,
Fig.~\ref{fig:adiabat}\,(c) shows that the conductivity decreases as
$T$ is lowered, with $d\sigma_{\rm dc}/dT > 0 $, indicating an
insulating behavior for all values of $\Delta$.
In line with this,
the pairing susceptibility, shown in Fig.\,\ref{fig:adiabat}\,(d),
remains small for all $\Delta$, suggesting that local electron pairs
are not correlated.  

\begin{figure}[t]
\hspace*{-0.3cm}
\includegraphics[scale=0.32]{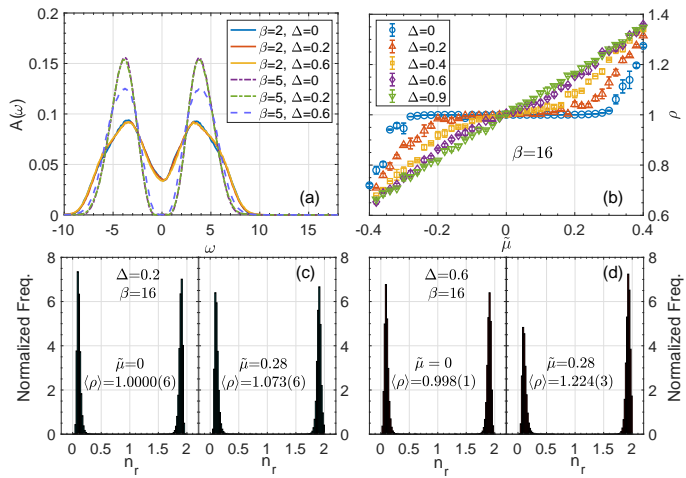}
\caption{The (a) density of states as a function of energy, (b)
  electron density, $\rho$, as a function of shifted chemical potential,
  $\tilde{\mu}=\mu+2g^{2}/\omega_{0}$, and the electron distribution
  at half-filling (Left) and (Right) away from half-filling at fixed
  (c) $\Delta=0.2$ and (d) $\Delta=0.6$.  $L=10$, $\omega=0.5$ and
  $\lambda_{D}=0.5$ ($g=1$).  }
\label{fig:CTAw} 
\end{figure}

We now characterize in more detail the large $\Delta$ behavior.
Figure~\ref{fig:CTAw}\,(a) shows the spectral function $A(\omega)$,
obtained via the analytic continuation of $G({\bf q},\tau) = \langle
\mathcal{T} d({\bf q},\tau) d^\dagger({\bf q},0)
\rangle=\int_{-\infty}^{\infty} d\omega \frac{e^{-\tau\omega}}
{1+e^{-\beta\omega}}A({\bf q},\omega)$, where
$\mathcal{T}$ is the imaginary time ordering operator, and
$A(\omega)$ sums over all momenta; see, e.g., the SM.
The spectral weight at the Fermi level is suppressed at low
$T$, with an opening of a single-particle gap.  This
occurs for both clean and disordered cases, even for large
disorder, where the CDW has been completely destroyed,
suggesting an insulating behavior for any disorder
strength.  Typically, the opening of such gaps in
$A(\omega)$ is associated with a vanishing compressibility
$\kappa=d\rho/d\mu$.  This happens, e.g., in the
half-filled fermionic Hubbard model, both in the
weak-coupling Slater and strong-coupling Mott regimes.
Similarly, in our disordered Holstein model the
compressibility also vanishes at weak disorder, as shown
in Fig.\,\ref{fig:CTAw}\,(b).  However, at large $\Delta$,
the gap in $A(\omega)$ is {\it not} accompanied by
$\kappa=0$.  As displayed in Fig.\,\ref{fig:CTAw}\,(b),
the plateau in $\rho(\mu)$ is substantially smeared at
$\Delta/t \sim 0.4$, and completely destroyed at $\Delta/t \sim 0.6$.

In both band and Mott insulators, $A(\omega)=0$
and $\kappa=0$ go hand-in-hand.
The unusual behavior 
whereby $A(\omega=0)=0$ but $\kappa \neq 0$
derives from the fact that the effective local
attractive interaction, due to phonon modes, favors the addition of
pairs of fermions to the system, while resisting the addition of
individual ones.  This picture is supported by analyzing the electron
distribution on the lattice during the Monte Carlo simulations.  In
Figs.\,\ref{fig:CTAw}\,(c)-(d), histograms of the local density
$n_\mathbf{r}$ are sharply peaked around 0 and 2 but not 1 for all disorder strengths,
indicating that we mostly have doubly occupied or empty sites.
Similar distributions are also observed away from half-filling.  For
instance, fixing $\tilde{\mu}=\mu+2g^{2}/\omega_{0} = 0.28$, and
comparing the electron distribution at $\Delta/t=0.2$ with
$\Delta/t=0.6$, the same chemical potential adds more pairs of
electrons into the system and causes a more distinguished imbalance
between empty and doubly occupied sites at larger disorder.  This
supports the picture that adding pairs of electrons is the mechanism
by which the system responds to increasing $\mu$.  Unlike the repulsive Hubbard
model, where the electron-electron interaction $U$ favors moment
formation (singly occupied sites) and the random site energies favor
pairs, here the electron-phonon interaction, $g$, and $\Delta$ both
promote binding.  Together, the properties shown in
Fig.\,\ref{fig:CTAw} point to an
insulating phase characterized by a gapless fermion pair excitation,
but a gapped spectrum for single particle ones.

\begin{figure}[t]
\includegraphics[scale=0.3]{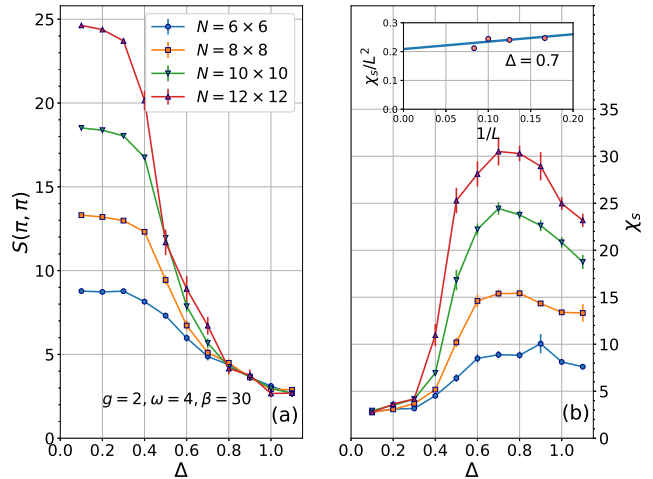}
\caption{The (a) charge structure factor $S(\pi,\pi)$, 
(b) $s$-wave pairing susceptibility $\chi_{s}$ as a function 
of disorder strength $\Delta$, at fixed $\beta=30$, $\omega_{0}=4$, 
and $\lambda_{D}=1/4$ ($g=2$).  Inset: The normalized pairing susceptibility $\chi_{s}/L^{2}$ 
as a function of $1/L$ at $\Delta=0.7$.} 
\label{fig:antiadiabat} 
\end{figure}

We now discuss the anti-adiabatic regime, fixing $\omega_0/t=4$ and
$\lambda_D=1/4$ ($g=2$).  Figure~\ref{fig:antiadiabat}\,(a) shows the
evolution of $S(\pi,\pi)$ with disorder, at a fixed low temperature
$T/t=1/30$.  As in the adiabatic regime, increasing $\Delta$ strongly
suppresses the charge response, destroying the CDW phase.  However, in
stark contrast with the former case, here the behavior of the pair
susceptibility is dramatically different: $\chi_{s}$ is two
orders of magnitude larger, and exhibits a peak around $\Delta/t=0.7$,
as displayed in Fig.\,\ref{fig:antiadiabat}\,(b).  
The magnitude of these charge structure
factors and superconducting susceptibilities are consistent with
those of their magnetic and pairing analogs indicating long range
order in the repulsive~\cite{varney09} and attractive Hubbard
models~\cite{scalettar89b,Paiva04,bouadim11}.
Although these large values of $\chi_{s}$ are
suggestive, finite size scaling (FSS) is required to establish the nature of the phase.
One approach to this FSS is to take data at very low temperatures, such as
$T/t=1/30$ in Fig.\,\ref{fig:antiadiabat} so that one is essentially at $T=0$,
on the simulated lattice size for that value of randomness.
The inset of Fig.\,\ref{fig:antiadiabat}\,(b) shows
that $\chi_{s}/L^{2}$, at $\Delta/t=0.7$, has a finite value
when extrapolated to $L\rightarrow\infty$, corresponding to long-range order and
a divergence of $\chi_{\rm
  pairing}$ in the thermodynamic limit. The qualitative picture is that, for these
parameters, disorder drives a SC state at commensurate filling as
charge correlations are suppressed, and new energy states are created
near the Fermi surface for pairing.
Given this, the results of these QMC
simulations is a crossover from a phase consisting of CDW-puddles to a
SC ordered one.

\begin{figure}[t]
\includegraphics[scale=0.25]{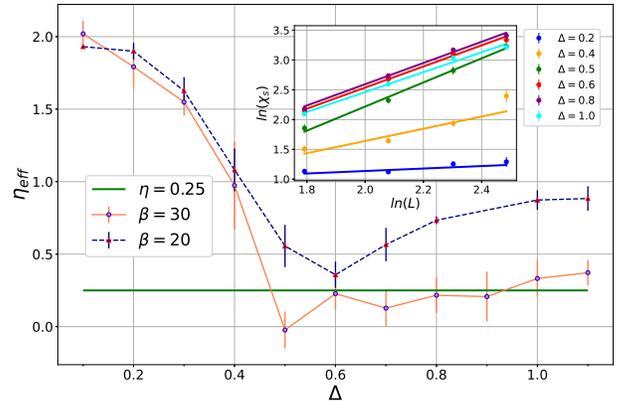}
\caption{The effective KT power law $\eta_{\rm eff}(T)$ is shown 
as a function of disorder $\Delta$ for two fixed low temperatures.  
$\eta_{\rm eff}(T) < 1/4$ for $T/t=1/30$ in a range of intermediate $\Delta$, 
suggesting a superconducting state.}
\label{fig:eta} 
\end{figure}

A more refined FSS analysis proceeds as follows:  
We expect the 2D superconducting transition
suggested by the data of Fig.~\ref{fig:antiadiabat} to be in 
the Kosterlitz-Thouless universality 
class.  Thus the pair susceptibility
$\chi_{s} \sim L^{2-\eta(T)}$ with a temperature-dependent exponent $\eta(T)$.
At the KT transition point $\eta(T_c) = 1/4$ and
$\eta(T) \rightarrow 0$ in the ground state.
Meanwhile, for $T>T_{KT}$, the pair correlations
decay exponentially on sufficiently large lattices,
 therefore $\chi_{s} \sim L^0$ according to
Eq.~\eqref{eq:chipairing}, i.e.~$\eta=2$.
Figure \ref{fig:eta} shows the results for such FSS analysis, 
in which we have used plots of
${\rm ln}(\chi_{s})$ versus ${\rm ln}(L)$ to extract $\eta_{\rm eff}$
at the fixed temperatures $T/t=1/20, 1/30$ of the simulations, 
as displayed in the inset.  We refer to this as an `effective' $\eta$ to acknowledge 
finite size effects.
The main panel of  Figure \ref{fig:eta} shows $\eta_{\rm eff}$ at these two temperatures
as a function of disorder $\Delta$.
At small $\Delta$, deep in the CDW phase, pairing correlations decay rapidly and we see the
expected $\eta_{\rm eff}=2$.
For $T/t = 1/20$, $\eta_{\rm eff}$ 
comes down rapidly as disorder strength is increased, indicative of pairing
correlations that are approaching the size of the lattice.
However, $\eta_{\rm eff}$ still exceeds
the universal KT value 
$\eta_{\rm eff}(T_c)=1/4$
for all $\Delta$.  There is no superconductivity at this temperature.
For $T/t = 1/30$, on the other hand, $\eta_{\rm eff} < 1/4$ in
a range of intermediate $\Delta$.  In this window, $T=1/30 < T_c$
and the system is in a superconducting phase.  The error bars are conservatively
estimated, and represent
a complex combination of statistical uncertainty for individual
disorder realizations, the disorder averaging, and
uncertainty associated with the FSS fit to 
extract $\eta$.


The overall picture which emerges from Figs.~\ref{fig:antiadiabat} and \ref{fig:eta} is that
substantial charge correlations are present at $T/t \lesssim 1/10$ in
the weak disorder region, $\Delta/t \lesssim 0.5$, while a SC dome
emerges for stronger disorder values at $T/t \lesssim 1/20$.  The issue
of how the CDW and SC phases meet at temperatures below $T=0.033$ is
beyond the scope of the present set of simulations.  
The heat maps of Fig.~S1 of
the SM suggest that there is a narrow region where both
$S(\pi,\pi)$ and $\chi_{s}$ are large.  However, while we
are able to perform definitive FSS analysis within the
individual CDW and SC phases, the corresponding data at the interface
between them do not provide an unambiguous conclusion. 
Furthermore, the coupling of random fields to
the CDW order parameter prevents the occurrence of true diagonal
long-range order~\cite{Vojta19}.  Notwithstanding, the emergence of SC is
allowed in the ground state, as indicated by 
our FSS analysis, and also emphasized in
the heat map presented in the SM.


\vskip0.02in \noindent
{\it Concluding Remarks.}
Although the two parameter regimes for which we have presented results
are distinguished by the value of $\omega_0/t$, we believe the
qualitative explanation for the difference in behavior, 
i.e.~the presence of an intermediate SC phase, lies
in the fact that the former corresponds to an intermediate and the latter to
a weak dimensionless coupling.  For strong and intermediate couplings, the
composite electron-phonon polarons are small, and hence easily
localized by disorder.  At weak dimensionless coupling, the polarons
are much larger, and the disorder potential is therefore to some
extent averaged out over their volume.  Thus, after $\Delta$ destroys
the CDW, it does not yet localize the pairs, which remain mobile and
condense into a SC phase.

Tuning between CDW and paired phases can be accomplished via pressure or
doping, and is a phenomenon which also has been extensively explored
experimentally.  Analogies between antiferromagnetic-SC and CDW-SC
phases have also been remarked~\cite{Scalettar89,esterlis18b}. However,
the latter transition has received much less attention from the QMC
community.  Early work on the doping-driven CDW-SC transition in the
Holstein model~\cite{Vekic92,Freericks93} has been extended to
transitions at commensurate filling caused by the introduction of band
dispersion~\cite{Costa18}, and a comparison with Migdal-Eliashberg
theory~\cite{esterlis18}.  Additional QMC literature has also considered 
the interplay between electron-electron and electron-phonon interactions, 
as in the Hubbard-Holstein
model~\cite{yamazaki14,Karakuzu17,Ohgoe17,Costa20,Wang19,Costa20b}.

This paper has described a detailed QMC study of the effect of disorder
on the CDW transition, and shown that, in certain parameter regimes,
randomness can give rise to a SC state.  Earlier work has suggested that
the electron-phonon coupling can renormalize the disorder potentials,
leading to a ground state that may not exhibit Anderson
localization~\cite{Bronold02,Ebrahimnejad12,Ebrahimnejad12b,Tozer14}.
The present study suggests an even more subtle consequence of the
disorder-interaction interplay, the emergence of a off-diagonal ordered
phases from diagonal disorder at commensurate filling.

We expect our results to apply quite generally to the 
Holstein model on other bipartite geometries (e.g.~3D cubic)
where CDW order is dominant at half-filling~\cite{Zhang19,Chen19,Costa20b}.  The honeycomb lattice
might be particularly interesting to investigate, 
since it has a quantum critical point for couplings below which CDW order is
absent.  SC might still emerge with added disorder in this semi-metallic regime
from the filling up of the density of states, which vanishes linearly in the clean limit.
We also expect our results to apply generally to different choices of $\lambda, \omega_0$
which have the same $\lambda_D$~\cite{zhang20}.  In the clean case,
the CDW transition temperature has recently been found as
as a function of $\lambda_D$~\cite{Weber18,Zhang19,Chen19},
a feature whose behavior with randomness would be interesting
to examine in future work.

\noindent
{\it Acknowledgements.}
The work of B.X.~and R.S.~was supported by the grant DE‐SC0014671 funded by
the US Department of Energy, Office of Science.
N.C.C.~was partially supported by the Brazilian
funding agencies CAPES and CNPq, and also acknowledges PRACE 
for awarding him access to Marconi at CINECA, Italy (PRACE-2019204934).
E.K.~acknowledges support from the 
National Science Foundation under grant No.~DMR-1918572.  G.G.B.
~acknowledges support from the University of the C\^ote d'Azur IDEX Jedi and Beijing CSRC.  
Computations were performed on the Spartan facility supported by NSF
OAC-162664 at SJSU.

\bibliography{bib_Hols}

\begin{thebibliography}{66}%
\makeatletter
\providecommand \@ifxundefined [1]{%
 \@ifx{#1\undefined}
}%
\providecommand \@ifnum [1]{%
 \ifnum #1\expandafter \@firstoftwo
 \else \expandafter \@secondoftwo
 \fi
}%
\providecommand \@ifx [1]{%
 \ifx #1\expandafter \@firstoftwo
 \else \expandafter \@secondoftwo
 \fi
}%
\providecommand \natexlab [1]{#1}%
\providecommand \enquote  [1]{``#1''}%
\providecommand \bibnamefont  [1]{#1}%
\providecommand \bibfnamefont [1]{#1}%
\providecommand \citenamefont [1]{#1}%
\providecommand \href@noop [0]{\@secondoftwo}%
\providecommand \href [0]{\begingroup \@sanitize@url \@href}%
\providecommand \@href[1]{\@@startlink{#1}\@@href}%
\providecommand \@@href[1]{\endgroup#1\@@endlink}%
\providecommand \@sanitize@url [0]{\catcode `\\12\catcode `\$12\catcode
  `\&12\catcode `\#12\catcode `\^12\catcode `\_12\catcode `\%12\relax}%
\providecommand \@@startlink[1]{}%
\providecommand \@@endlink[0]{}%
\providecommand \url  [0]{\begingroup\@sanitize@url \@url }%
\providecommand \@url [1]{\endgroup\@href {#1}{\urlprefix }}%
\providecommand \urlprefix  [0]{URL }%
\providecommand \Eprint [0]{\href }%
\providecommand \doibase [0]{http://dx.doi.org/}%
\providecommand \selectlanguage [0]{\@gobble}%
\providecommand \bibinfo  [0]{\@secondoftwo}%
\providecommand \bibfield  [0]{\@secondoftwo}%
\providecommand \translation [1]{[#1]}%
\providecommand \BibitemOpen [0]{}%
\providecommand \bibitemStop [0]{}%
\providecommand \bibitemNoStop [0]{.\EOS\space}%
\providecommand \EOS [0]{\spacefactor3000\relax}%
\providecommand \BibitemShut  [1]{\csname bibitem#1\endcsname}%
\let\auto@bib@innerbib\@empty
\bibitem [{\citenamefont {Abrahams}\ \emph {et~al.}(1979)\citenamefont
  {Abrahams}, \citenamefont {Anderson}, \citenamefont {Licciardello},\ and\
  \citenamefont {Ramakrishnan}}]{abrahams79}%
  \BibitemOpen
  \bibfield  {author} {\bibinfo {author} {\bibfnamefont {Elihu}\ \bibnamefont
  {Abrahams}}, \bibinfo {author} {\bibfnamefont {PW}~\bibnamefont {Anderson}},
  \bibinfo {author} {\bibfnamefont {DC}~\bibnamefont {Licciardello}}, \ and\
  \bibinfo {author} {\bibfnamefont {TV}~\bibnamefont {Ramakrishnan}},\
  }\bibfield  {title} {\enquote {\bibinfo {title} {Scaling theory of
  localization: Absence of quantum diffusion in two dimensions},}\ }\href@noop
  {} {\bibfield  {journal} {\bibinfo  {journal} {Phys. Rev. Lett.}\ }\textbf
  {\bibinfo {volume} {42}},\ \bibinfo {pages} {673} (\bibinfo {year}
  {1979})}\BibitemShut {NoStop}%
\bibitem [{\citenamefont {Wegner}(1980)}]{wegner80}%
  \BibitemOpen
  \bibfield  {author} {\bibinfo {author} {\bibfnamefont {Franz}\ \bibnamefont
  {Wegner}},\ }\bibfield  {title} {\enquote {\bibinfo {title} {Inverse
  participation ratio in 2+ $\varepsilon$ dimensions},}\ }\href@noop {}
  {\bibfield  {journal} {\bibinfo  {journal} {Zeitschrift f{\"u}r Physik B
  Condensed Matter}\ }\textbf {\bibinfo {volume} {36}},\ \bibinfo {pages}
  {209--214} (\bibinfo {year} {1980})}\BibitemShut {NoStop}%
\bibitem [{\citenamefont {Efetov}\ \emph {et~al.}(1980)\citenamefont {Efetov},
  \citenamefont {Larkin},\ and\ \citenamefont {Khmelnitskii}}]{efetov80}%
  \BibitemOpen
  \bibfield  {author} {\bibinfo {author} {\bibfnamefont {K.B.}\ \bibnamefont
  {Efetov}}, \bibinfo {author} {\bibfnamefont {A.I.}\ \bibnamefont {Larkin}}, \
  and\ \bibinfo {author} {\bibfnamefont {D.~E.}\ \bibnamefont {Khmelnitskii}},\
  }\href@noop {} {\bibfield  {journal} {\bibinfo  {journal} {Sov. Phys. JETP}\
  }\textbf {\bibinfo {volume} {52}},\ \bibinfo {pages} {568} (\bibinfo {year}
  {1980})}\BibitemShut {NoStop}%
\bibitem [{\citenamefont {Lee}\ and\ \citenamefont
  {Ramakrishnan}(1985)}]{lee85}%
  \BibitemOpen
  \bibfield  {author} {\bibinfo {author} {\bibfnamefont {Patrick~A.}\
  \bibnamefont {Lee}}\ and\ \bibinfo {author} {\bibfnamefont {T.~V.}\
  \bibnamefont {Ramakrishnan}},\ }\bibfield  {title} {\enquote {\bibinfo
  {title} {Disordered electronic systems},}\ }\href {\doibase
  10.1103/RevModPhys.57.287} {\bibfield  {journal} {\bibinfo  {journal} {Rev.
  Mod. Phys.}\ }\textbf {\bibinfo {volume} {57}},\ \bibinfo {pages} {287--337}
  (\bibinfo {year} {1985})}\BibitemShut {NoStop}%
\bibitem [{\citenamefont {Giamarchi}\ and\ \citenamefont
  {Schulz}(1988)}]{giamarchi88}%
  \BibitemOpen
  \bibfield  {author} {\bibinfo {author} {\bibfnamefont {T.}~\bibnamefont
  {Giamarchi}}\ and\ \bibinfo {author} {\bibfnamefont {H.~J.}\ \bibnamefont
  {Schulz}},\ }\bibfield  {title} {\enquote {\bibinfo {title} {Anderson
  localization and interactions in one-dimensional metals},}\ }\href {\doibase
  10.1103/PhysRevB.37.325} {\bibfield  {journal} {\bibinfo  {journal} {Phys.
  Rev. B}\ }\textbf {\bibinfo {volume} {37}},\ \bibinfo {pages} {325--340}
  (\bibinfo {year} {1988})}\BibitemShut {NoStop}%
\bibitem [{\citenamefont {Belitz}\ and\ \citenamefont
  {Kirkpatrick}(1994)}]{belitz94}%
  \BibitemOpen
  \bibfield  {author} {\bibinfo {author} {\bibfnamefont {D.}~\bibnamefont
  {Belitz}}\ and\ \bibinfo {author} {\bibfnamefont {T.~R.}\ \bibnamefont
  {Kirkpatrick}},\ }\bibfield  {title} {\enquote {\bibinfo {title} {{The
  Anderson-Mott transition}},}\ }\href {\doibase 10.1103/RevModPhys.66.261}
  {\bibfield  {journal} {\bibinfo  {journal} {Rev. Mod. Phys.}\ }\textbf
  {\bibinfo {volume} {66}},\ \bibinfo {pages} {261--380} (\bibinfo {year}
  {1994})}\BibitemShut {NoStop}%
\bibitem [{\citenamefont {Dagotto}(2005)}]{dagotto05}%
  \BibitemOpen
  \bibfield  {author} {\bibinfo {author} {\bibfnamefont {Elbio}\ \bibnamefont
  {Dagotto}},\ }\bibfield  {title} {\enquote {\bibinfo {title} {Complexity in
  strongly correlated electronic systems},}\ }\href {\doibase
  10.1126/science.1107559} {\bibfield  {journal} {\bibinfo  {journal}
  {Science}\ }\textbf {\bibinfo {volume} {309}},\ \bibinfo {pages} {257--262}
  (\bibinfo {year} {2005})}\BibitemShut {NoStop}%
\bibitem [{\citenamefont {Balatsky}\ \emph {et~al.}(2006)\citenamefont
  {Balatsky}, \citenamefont {Vekhter},\ and\ \citenamefont {Zhu}}]{balatsky06}%
  \BibitemOpen
  \bibfield  {author} {\bibinfo {author} {\bibfnamefont {A.~V.}\ \bibnamefont
  {Balatsky}}, \bibinfo {author} {\bibfnamefont {I.}~\bibnamefont {Vekhter}}, \
  and\ \bibinfo {author} {\bibfnamefont {Jian-Xin}\ \bibnamefont {Zhu}},\
  }\bibfield  {title} {\enquote {\bibinfo {title} {Impurity-induced states in
  conventional and unconventional superconductors},}\ }\href {\doibase
  10.1103/RevModPhys.78.373} {\bibfield  {journal} {\bibinfo  {journal} {Rev.
  Mod. Phys.}\ }\textbf {\bibinfo {volume} {78}},\ \bibinfo {pages} {373--433}
  (\bibinfo {year} {2006})}\BibitemShut {NoStop}%
\bibitem [{\citenamefont {Abrahams}(2010)}]{abrahams10}%
  \BibitemOpen
  \bibfield  {author} {\bibinfo {author} {\bibfnamefont {E.}~\bibnamefont
  {Abrahams}},\ }\href@noop {} {\emph {\bibinfo {title} {50 Years of Anderson
  Localization}}}\ (\bibinfo  {publisher} {World Scientific},\ \bibinfo {year}
  {2010})\BibitemShut {NoStop}%
\bibitem [{\citenamefont {Dobrosavljevic}\ \emph {et~al.}(2012)\citenamefont
  {Dobrosavljevic}, \citenamefont {Trivedi},\ and\ \citenamefont
  {Valles~Jr}}]{dobrosavljevic12}%
  \BibitemOpen
  \bibfield  {author} {\bibinfo {author} {\bibfnamefont {V.}~\bibnamefont
  {Dobrosavljevic}}, \bibinfo {author} {\bibfnamefont {N.}~\bibnamefont
  {Trivedi}}, \ and\ \bibinfo {author} {\bibfnamefont {James~M}\ \bibnamefont
  {Valles~Jr}},\ }\href@noop {} {\emph {\bibinfo {title} {Conductor-insulator
  quantum phase transitions}}}\ (\bibinfo  {publisher} {Oxford University
  Press, Oxford, UK},\ \bibinfo {year} {2012})\BibitemShut {NoStop}%
\bibitem [{\citenamefont {Vojta}(2019)}]{Vojta19}%
  \BibitemOpen
  \bibfield  {author} {\bibinfo {author} {\bibfnamefont {T.}~\bibnamefont
  {Vojta}},\ }\bibfield  {title} {\enquote {\bibinfo {title} {Disorder in
  quantum many-body systems},}\ }\href {\doibase
  10.1146/annurev-conmatphys-031218-013433} {\bibfield  {journal} {\bibinfo
  {journal} {Annual Review of Condensed Matter Physics}\ }\textbf {\bibinfo
  {volume} {10}},\ \bibinfo {pages} {233--252} (\bibinfo {year}
  {2019})}\BibitemShut {NoStop}%
\bibitem [{\citenamefont {Kravchenko}\ and\ \citenamefont
  {Sarachik}(2003)}]{Kravchenko03}%
  \BibitemOpen
  \bibfield  {author} {\bibinfo {author} {\bibfnamefont {S.V.}\ \bibnamefont
  {Kravchenko}}\ and\ \bibinfo {author} {\bibfnamefont {M.P.}\ \bibnamefont
  {Sarachik}},\ }\bibfield  {title} {\enquote {\bibinfo {title}
  {Metal{\textendash}insulator transition in two-dimensional electron
  systems},}\ }\href {\doibase 10.1088/0034-4885/67/1/r01} {\bibfield
  {journal} {\bibinfo  {journal} {Reports on Progress in Physics}\ }\textbf
  {\bibinfo {volume} {67}},\ \bibinfo {pages} {1--44} (\bibinfo {year}
  {2003})}\BibitemShut {NoStop}%
\bibitem [{\citenamefont {Lobos}\ \emph {et~al.}(2012)\citenamefont {Lobos},
  \citenamefont {Lutchyn},\ and\ \citenamefont {Das~Sarma}}]{lobos12}%
  \BibitemOpen
  \bibfield  {author} {\bibinfo {author} {\bibfnamefont {Alejandro~M.}\
  \bibnamefont {Lobos}}, \bibinfo {author} {\bibfnamefont {Roman~M.}\
  \bibnamefont {Lutchyn}}, \ and\ \bibinfo {author} {\bibfnamefont
  {S.}~\bibnamefont {Das~Sarma}},\ }\bibfield  {title} {\enquote {\bibinfo
  {title} {Interplay of disorder and interaction in majorana quantum wires},}\
  }\href {\doibase 10.1103/PhysRevLett.109.146403} {\bibfield  {journal}
  {\bibinfo  {journal} {Phys. Rev. Lett.}\ }\textbf {\bibinfo {volume} {109}},\
  \bibinfo {pages} {146403} (\bibinfo {year} {2012})}\BibitemShut {NoStop}%
\bibitem [{\citenamefont {Krishna}\ \emph {et~al.}(2019)\citenamefont
  {Krishna}, \citenamefont {Ippoliti},\ and\ \citenamefont
  {Bhatt}}]{krishna19}%
  \BibitemOpen
  \bibfield  {author} {\bibinfo {author} {\bibfnamefont {Akshay}\ \bibnamefont
  {Krishna}}, \bibinfo {author} {\bibfnamefont {Matteo}\ \bibnamefont
  {Ippoliti}}, \ and\ \bibinfo {author} {\bibfnamefont {R.~N.}\ \bibnamefont
  {Bhatt}},\ }\bibfield  {title} {\enquote {\bibinfo {title} {Localization and
  interactions in topological and nontopological bands in two dimensions},}\
  }\href {\doibase 10.1103/PhysRevB.100.054202} {\bibfield  {journal} {\bibinfo
   {journal} {Phys. Rev. B}\ }\textbf {\bibinfo {volume} {100}},\ \bibinfo
  {pages} {054202} (\bibinfo {year} {2019})}\BibitemShut {NoStop}%
\bibitem [{\citenamefont {Kondov}\ \emph {et~al.}(2011)\citenamefont {Kondov},
  \citenamefont {McGehee}, \citenamefont {Zirbel},\ and\ \citenamefont
  {DeMarco}}]{kondov11}%
  \BibitemOpen
  \bibfield  {author} {\bibinfo {author} {\bibfnamefont {S.~S.}\ \bibnamefont
  {Kondov}}, \bibinfo {author} {\bibfnamefont {W.~R.}\ \bibnamefont {McGehee}},
  \bibinfo {author} {\bibfnamefont {J.~J.}\ \bibnamefont {Zirbel}}, \ and\
  \bibinfo {author} {\bibfnamefont {B.}~\bibnamefont {DeMarco}},\ }\bibfield
  {title} {\enquote {\bibinfo {title} {Three-dimensional anderson localization
  of ultracold matter},}\ }\href {\doibase 10.1126/science.1209019} {\bibfield
  {journal} {\bibinfo  {journal} {Science}\ }\textbf {\bibinfo {volume}
  {334}},\ \bibinfo {pages} {66--68} (\bibinfo {year} {2011})}\BibitemShut
  {NoStop}%
\bibitem [{\citenamefont {Basko}\ \emph {et~al.}(2006)\citenamefont {Basko},
  \citenamefont {Aleiner},\ and\ \citenamefont {Altshuler}}]{basko06}%
  \BibitemOpen
  \bibfield  {author} {\bibinfo {author} {\bibfnamefont {D.~M.}\ \bibnamefont
  {Basko}}, \bibinfo {author} {\bibfnamefont {I.~L.}\ \bibnamefont {Aleiner}},
  \ and\ \bibinfo {author} {\bibfnamefont {B.~L.}\ \bibnamefont {Altshuler}},\
  }\href@noop {} {\bibfield  {journal} {\bibinfo  {journal} {Ann. Phys. (NY)}\
  }\textbf {\bibinfo {volume} {321}},\ \bibinfo {pages} {1126} (\bibinfo {year}
  {2006})}\BibitemShut {NoStop}%
\bibitem [{\citenamefont {Nandkishore}\ and\ \citenamefont
  {Huse}(2015)}]{nandkishore15}%
  \BibitemOpen
  \bibfield  {author} {\bibinfo {author} {\bibfnamefont {R.}~\bibnamefont
  {Nandkishore}}\ and\ \bibinfo {author} {\bibfnamefont {D.~A.}\ \bibnamefont
  {Huse}},\ }\href@noop {} {\bibfield  {journal} {\bibinfo  {journal} {Annu.
  Rev. Condens. Matter Phys.}\ }\textbf {\bibinfo {volume} {6}},\ \bibinfo
  {pages} {15} (\bibinfo {year} {2015})}\BibitemShut {NoStop}%
\bibitem [{\citenamefont {Lev}\ \emph {et~al.}(2015)\citenamefont {Lev},
  \citenamefont {Cohen},\ and\ \citenamefont {Reichman}}]{barlev15}%
  \BibitemOpen
  \bibfield  {author} {\bibinfo {author} {\bibfnamefont {Yevgeny~Bar}\
  \bibnamefont {Lev}}, \bibinfo {author} {\bibfnamefont {Guy}\ \bibnamefont
  {Cohen}}, \ and\ \bibinfo {author} {\bibfnamefont {David~R}\ \bibnamefont
  {Reichman}},\ }\bibfield  {title} {\enquote {\bibinfo {title} {Absence of
  diffusion in an interacting system of spinless fermions on a one-dimensional
  disordered lattice},}\ }\href@noop {} {\bibfield  {journal} {\bibinfo
  {journal} {Phys. Rev. Lett.}\ }\textbf {\bibinfo {volume} {114}},\ \bibinfo
  {pages} {100601} (\bibinfo {year} {2015})}\BibitemShut {NoStop}%
\bibitem [{\citenamefont {Denteneer}\ \emph {et~al.}(1999)\citenamefont
  {Denteneer}, \citenamefont {Scalettar},\ and\ \citenamefont
  {Trivedi}}]{Denteneer99}%
  \BibitemOpen
  \bibfield  {author} {\bibinfo {author} {\bibfnamefont {P.~J.~H.}\
  \bibnamefont {Denteneer}}, \bibinfo {author} {\bibfnamefont {R.~T.}\
  \bibnamefont {Scalettar}}, \ and\ \bibinfo {author} {\bibfnamefont
  {N.}~\bibnamefont {Trivedi}},\ }\bibfield  {title} {\enquote {\bibinfo
  {title} {{Conducting phase in the two-dimensional disordered Hubbard
  model}},}\ }\href {\doibase 10.1103/PhysRevLett.83.4610} {\bibfield
  {journal} {\bibinfo  {journal} {Phys. Rev. Lett.}\ }\textbf {\bibinfo
  {volume} {83}},\ \bibinfo {pages} {4610--4613} (\bibinfo {year}
  {1999})}\BibitemShut {NoStop}%
\bibitem [{\citenamefont {Terletska}\ \emph {et~al.}(2018)\citenamefont
  {Terletska}, \citenamefont {Zhang}, \citenamefont {Tam}, \citenamefont
  {Berlijn}, \citenamefont {Chioncel}, \citenamefont {Vidhyadhiraja},\ and\
  \citenamefont {Jarrell}}]{terletska18}%
  \BibitemOpen
  \bibfield  {author} {\bibinfo {author} {\bibfnamefont {H.}~\bibnamefont
  {Terletska}}, \bibinfo {author} {\bibfnamefont {Y.}~\bibnamefont {Zhang}},
  \bibinfo {author} {\bibfnamefont {K-M.}\ \bibnamefont {Tam}}, \bibinfo
  {author} {\bibfnamefont {T.}~\bibnamefont {Berlijn}}, \bibinfo {author}
  {\bibfnamefont {L.}~\bibnamefont {Chioncel}}, \bibinfo {author}
  {\bibfnamefont {N.S.}\ \bibnamefont {Vidhyadhiraja}}, \ and\ \bibinfo
  {author} {\bibfnamefont {M.}~\bibnamefont {Jarrell}},\ }\bibfield  {title}
  {\enquote {\bibinfo {title} {Systematic quantum cluster typical medium method
  for the study of localization in strongly disordered electronic systems},}\
  }\href@noop {} {\bibfield  {journal} {\bibinfo  {journal} {Appl. Sci.}\
  }\textbf {\bibinfo {volume} {8}},\ \bibinfo {pages} {2401} (\bibinfo {year}
  {2018})}\BibitemShut {NoStop}%
\bibitem [{\citenamefont {Loh}\ \emph {et~al.}(1990)\citenamefont {Loh},
  \citenamefont {Gubernatis}, \citenamefont {Scalettar}, \citenamefont {White},
  \citenamefont {Scalapino},\ and\ \citenamefont {Sugar}}]{loh90}%
  \BibitemOpen
  \bibfield  {author} {\bibinfo {author} {\bibfnamefont {E.~Y.}\ \bibnamefont
  {Loh}}, \bibinfo {author} {\bibfnamefont {J.~E.}\ \bibnamefont {Gubernatis}},
  \bibinfo {author} {\bibfnamefont {R.~T.}\ \bibnamefont {Scalettar}}, \bibinfo
  {author} {\bibfnamefont {S.~R.}\ \bibnamefont {White}}, \bibinfo {author}
  {\bibfnamefont {D.~J.}\ \bibnamefont {Scalapino}}, \ and\ \bibinfo {author}
  {\bibfnamefont {R.~L.}\ \bibnamefont {Sugar}},\ }\bibfield  {title} {\enquote
  {\bibinfo {title} {Sign problem in the numerical simulation of many-electron
  systems},}\ }\href {\doibase 10.1103/PhysRevB.41.9301} {\bibfield  {journal}
  {\bibinfo  {journal} {Phys. Rev. B}\ }\textbf {\bibinfo {volume} {41}},\
  \bibinfo {pages} {9301--9307} (\bibinfo {year} {1990})}\BibitemShut {NoStop}%
\bibitem [{\citenamefont {Troyer}\ and\ \citenamefont
  {Wiese}(2005)}]{troyer05}%
  \BibitemOpen
  \bibfield  {author} {\bibinfo {author} {\bibfnamefont {M.}~\bibnamefont
  {Troyer}}\ and\ \bibinfo {author} {\bibfnamefont {U-J.}\ \bibnamefont
  {Wiese}},\ }\bibfield  {title} {\enquote {\bibinfo {title} {Computational
  complexity and fundamental limitations to fermionic quantum monte carlo
  simulations},}\ }\href {\doibase 10.1103/PhysRevLett.94.170201} {\bibfield
  {journal} {\bibinfo  {journal} {Phys. Rev. Lett.}\ }\textbf {\bibinfo
  {volume} {94}},\ \bibinfo {pages} {170201} (\bibinfo {year}
  {2005})}\BibitemShut {NoStop}%
\bibitem [{\citenamefont {Iglovikov}\ \emph {et~al.}(2015)\citenamefont
  {Iglovikov}, \citenamefont {Khatami},\ and\ \citenamefont
  {Scalettar}}]{iglovikov15}%
  \BibitemOpen
  \bibfield  {author} {\bibinfo {author} {\bibfnamefont {V.~I.}\ \bibnamefont
  {Iglovikov}}, \bibinfo {author} {\bibfnamefont {E.}~\bibnamefont {Khatami}},
  \ and\ \bibinfo {author} {\bibfnamefont {R.~T.}\ \bibnamefont {Scalettar}},\
  }\bibfield  {title} {\enquote {\bibinfo {title} {{Geometry dependence of the
  sign problem in quantum Monte Carlo simulations}},}\ }\href {\doibase
  10.1103/PhysRevB.92.045110} {\bibfield  {journal} {\bibinfo  {journal} {Phys.
  Rev. B}\ }\textbf {\bibinfo {volume} {92}},\ \bibinfo {pages} {045110}
  (\bibinfo {year} {2015})}\BibitemShut {NoStop}%
\bibitem [{\citenamefont {Weber}\ and\ \citenamefont
  {Hohenadler}(2018)}]{Weber18}%
  \BibitemOpen
  \bibfield  {author} {\bibinfo {author} {\bibfnamefont {Manuel}\ \bibnamefont
  {Weber}}\ and\ \bibinfo {author} {\bibfnamefont {Martin}\ \bibnamefont
  {Hohenadler}},\ }\bibfield  {title} {\enquote {\bibinfo {title}
  {{Two-dimensional Holstein-Hubbard model: Critical temperature, Ising
  universality, and bipolaron liquid}},}\ }\href {\doibase
  10.1103/PhysRevB.98.085405} {\bibfield  {journal} {\bibinfo  {journal} {Phys.
  Rev. B}\ }\textbf {\bibinfo {volume} {98}},\ \bibinfo {pages} {085405}
  (\bibinfo {year} {2018})}\BibitemShut {NoStop}%
\bibitem [{\citenamefont {Zhang}\ \emph {et~al.}(2019)\citenamefont {Zhang},
  \citenamefont {Chiu}, \citenamefont {Costa}, \citenamefont {Batrouni},\ and\
  \citenamefont {Scalettar}}]{Zhang19}%
  \BibitemOpen
  \bibfield  {author} {\bibinfo {author} {\bibfnamefont {Y.-X.}\ \bibnamefont
  {Zhang}}, \bibinfo {author} {\bibfnamefont {W.-T.}\ \bibnamefont {Chiu}},
  \bibinfo {author} {\bibfnamefont {N.~C.}\ \bibnamefont {Costa}}, \bibinfo
  {author} {\bibfnamefont {G.~G.}\ \bibnamefont {Batrouni}}, \ and\ \bibinfo
  {author} {\bibfnamefont {R.~T.}\ \bibnamefont {Scalettar}},\ }\bibfield
  {title} {\enquote {\bibinfo {title} {{Charge order in the Holstein model on a
  honeycomb lattice}},}\ }\href {\doibase 10.1103/PhysRevLett.122.077602}
  {\bibfield  {journal} {\bibinfo  {journal} {Phys. Rev. Lett.}\ }\textbf
  {\bibinfo {volume} {122}},\ \bibinfo {pages} {077602} (\bibinfo {year}
  {2019})}\BibitemShut {NoStop}%
\bibitem [{\citenamefont {Chen}\ \emph {et~al.}(2019)\citenamefont {Chen},
  \citenamefont {Xu}, \citenamefont {Meng},\ and\ \citenamefont
  {Hohenadler}}]{Chen19}%
  \BibitemOpen
  \bibfield  {author} {\bibinfo {author} {\bibfnamefont {Chuang}\ \bibnamefont
  {Chen}}, \bibinfo {author} {\bibfnamefont {Xiao~Yan}\ \bibnamefont {Xu}},
  \bibinfo {author} {\bibfnamefont {Zi~Yang}\ \bibnamefont {Meng}}, \ and\
  \bibinfo {author} {\bibfnamefont {Martin}\ \bibnamefont {Hohenadler}},\
  }\bibfield  {title} {\enquote {\bibinfo {title} {{Charge-density-wave
  transitions of Dirac fermions coupled to phonons}},}\ }\href {\doibase
  10.1103/PhysRevLett.122.077601} {\bibfield  {journal} {\bibinfo  {journal}
  {Phys. Rev. Lett.}\ }\textbf {\bibinfo {volume} {122}},\ \bibinfo {pages}
  {077601} (\bibinfo {year} {2019})}\BibitemShut {NoStop}%
\bibitem [{\citenamefont {Cohen-Stead}\ \emph {et~al.}(2020)\citenamefont
  {Cohen-Stead}, \citenamefont {Barros}, \citenamefont {Meng}, \citenamefont
  {Chen}, \citenamefont {Scalettar},\ and\ \citenamefont
  {Batrouni}}]{CohenStead20}%
  \BibitemOpen
  \bibfield  {author} {\bibinfo {author} {\bibfnamefont {B.}~\bibnamefont
  {Cohen-Stead}}, \bibinfo {author} {\bibfnamefont {Kipton}\ \bibnamefont
  {Barros}}, \bibinfo {author} {\bibfnamefont {ZY}~\bibnamefont {Meng}},
  \bibinfo {author} {\bibfnamefont {Chuang}\ \bibnamefont {Chen}}, \bibinfo
  {author} {\bibfnamefont {R.~T.}\ \bibnamefont {Scalettar}}, \ and\ \bibinfo
  {author} {\bibfnamefont {G.~G.}\ \bibnamefont {Batrouni}},\ }\bibfield
  {title} {\enquote {\bibinfo {title} {Langevin simulations of the half-filled
  cubic {H}olstein model},}\ }\href {\doibase 10.1103/PhysRevB.102.161108}
  {\bibfield  {journal} {\bibinfo  {journal} {Phys. Rev. B}\ }\textbf {\bibinfo
  {volume} {102}},\ \bibinfo {pages} {161108} (\bibinfo {year}
  {2020})}\BibitemShut {NoStop}%
\bibitem [{\citenamefont {Blankenbecler}\ \emph {et~al.}(1981)\citenamefont
  {Blankenbecler}, \citenamefont {Scalapino},\ and\ \citenamefont
  {Sugar}}]{Blankenbecler81}%
  \BibitemOpen
  \bibfield  {author} {\bibinfo {author} {\bibfnamefont {R.}~\bibnamefont
  {Blankenbecler}}, \bibinfo {author} {\bibfnamefont {D.~J.}\ \bibnamefont
  {Scalapino}}, \ and\ \bibinfo {author} {\bibfnamefont {R.~L.}\ \bibnamefont
  {Sugar}},\ }\bibfield  {title} {\enquote {\bibinfo {title} {{Monte Carlo
  calculations of coupled boson-fermion systems. I}},}\ }\href {\doibase
  10.1103/PhysRevD.24.2278} {\bibfield  {journal} {\bibinfo  {journal} {Phys.
  Rev. D}\ }\textbf {\bibinfo {volume} {24}},\ \bibinfo {pages} {2278--2286}
  (\bibinfo {year} {1981})}\BibitemShut {NoStop}%
\bibitem [{\citenamefont {Scalettar}\ \emph
  {et~al.}(1989{\natexlab{a}})\citenamefont {Scalettar}, \citenamefont
  {Bickers},\ and\ \citenamefont {Scalapino}}]{Scalettar89}%
  \BibitemOpen
  \bibfield  {author} {\bibinfo {author} {\bibfnamefont {R.~T.}\ \bibnamefont
  {Scalettar}}, \bibinfo {author} {\bibfnamefont {N.~E.}\ \bibnamefont
  {Bickers}}, \ and\ \bibinfo {author} {\bibfnamefont {D.~J.}\ \bibnamefont
  {Scalapino}},\ }\bibfield  {title} {\enquote {\bibinfo {title} {{Competition
  of pairing and Peierls--charge-density-wave correlations in a two-dimensional
  electron-phonon model}},}\ }\href {\doibase 10.1103/PhysRevB.40.197}
  {\bibfield  {journal} {\bibinfo  {journal} {Phys. Rev. B}\ }\textbf {\bibinfo
  {volume} {40}},\ \bibinfo {pages} {197--200} (\bibinfo {year}
  {1989}{\natexlab{a}})}\BibitemShut {NoStop}%
\bibitem [{\citenamefont {Noack}\ \emph {et~al.}(1991)\citenamefont {Noack},
  \citenamefont {Scalapino},\ and\ \citenamefont {Scalettar}}]{Noack91}%
  \BibitemOpen
  \bibfield  {author} {\bibinfo {author} {\bibfnamefont {R.~M.}\ \bibnamefont
  {Noack}}, \bibinfo {author} {\bibfnamefont {D.~J.}\ \bibnamefont
  {Scalapino}}, \ and\ \bibinfo {author} {\bibfnamefont {R.~T.}\ \bibnamefont
  {Scalettar}},\ }\bibfield  {title} {\enquote {\bibinfo {title}
  {Charge-density-wave and pairing susceptibilities in a two-dimensional
  electron-phonon model},}\ }\href {\doibase 10.1103/PhysRevLett.66.778}
  {\bibfield  {journal} {\bibinfo  {journal} {Phys. Rev. Lett.}\ }\textbf
  {\bibinfo {volume} {66}},\ \bibinfo {pages} {778--781} (\bibinfo {year}
  {1991})}\BibitemShut {NoStop}%
\bibitem [{\citenamefont {\surname{dos Santos}}(2003)}]{Santos03}%
  \BibitemOpen
  \bibfield  {author} {\bibinfo {author} {\bibfnamefont {R.R.}\ \bibnamefont
  {\surname{dos Santos}}},\ }\bibfield  {title} {\enquote {\bibinfo {title}
  {{Introduction to quantum Monte Carlo simulations for fermionic systems}},}\
  }\href {\doibase 10.1590/S0103-97332003000100003} {\bibfield  {journal}
  {\bibinfo  {journal} {Brazilian Journal of Physics}\ }\textbf {\bibinfo
  {volume} {33}},\ \bibinfo {pages} {36 -- 54} (\bibinfo {year}
  {2003})}\BibitemShut {NoStop}%
\bibitem [{sup()}]{supplemental}%
  \BibitemOpen
  \href {https://journals.aps.org/supplemental/10.1103/PhysRevB.103.L060501}
  {}\bibinfo {note} {See Supplemental Material for more details about the
  CDW-SC competition in the anti-adiabatic limit, the Determinant Quantum Monte
  Carlo method, CDW transition in the clean limit, specific heat, temperature
  dependence in the anti-adiabatic limit, and disorder dependence in the
  adiabatic limit.}\BibitemShut {Stop}%
\bibitem [{\citenamefont {Huscroft}\ and\ \citenamefont
  {Scalettar}(1997)}]{Huscroft97}%
  \BibitemOpen
  \bibfield  {author} {\bibinfo {author} {\bibfnamefont {C.}~\bibnamefont
  {Huscroft}}\ and\ \bibinfo {author} {\bibfnamefont {R.~T.}\ \bibnamefont
  {Scalettar}},\ }\bibfield  {title} {\enquote {\bibinfo {title} {{Effect of
  disorder on charge-density wave and superconducting order in the half-filled
  attractive Hubbard model}},}\ }\href {\doibase 10.1103/PhysRevB.55.1185}
  {\bibfield  {journal} {\bibinfo  {journal} {Phys. Rev. B}\ }\textbf {\bibinfo
  {volume} {55}},\ \bibinfo {pages} {1185--1193} (\bibinfo {year}
  {1997})}\BibitemShut {NoStop}%
\bibitem [{\citenamefont {Paiva}\ \emph {et~al.}(2004)\citenamefont {Paiva},
  \citenamefont {dos Santos}, \citenamefont {Scalettar},\ and\ \citenamefont
  {Denteneer}}]{Paiva04}%
  \BibitemOpen
  \bibfield  {author} {\bibinfo {author} {\bibfnamefont {T.}~\bibnamefont
  {Paiva}}, \bibinfo {author} {\bibfnamefont {R.R.}\ \bibnamefont {dos
  Santos}}, \bibinfo {author} {\bibfnamefont {R.T.}\ \bibnamefont {Scalettar}},
  \ and\ \bibinfo {author} {\bibfnamefont {P.~J.~H.}\ \bibnamefont
  {Denteneer}},\ }\bibfield  {title} {\enquote {\bibinfo {title} {{Critical
  temperature for the two-dimensional attractive Hubbard model}},}\ }\href
  {\doibase 10.1103/PhysRevB.69.184501} {\bibfield  {journal} {\bibinfo
  {journal} {Phys. Rev. B}\ }\textbf {\bibinfo {volume} {69}},\ \bibinfo
  {pages} {184501} (\bibinfo {year} {2004})}\BibitemShut {NoStop}%
\bibitem [{\citenamefont {Trivedi}\ \emph {et~al.}(1996)\citenamefont
  {Trivedi}, \citenamefont {Scalettar},\ and\ \citenamefont
  {Randeria}}]{Trivedi96}%
  \BibitemOpen
  \bibfield  {author} {\bibinfo {author} {\bibfnamefont {Nandini}\ \bibnamefont
  {Trivedi}}, \bibinfo {author} {\bibfnamefont {Richard~T.}\ \bibnamefont
  {Scalettar}}, \ and\ \bibinfo {author} {\bibfnamefont {Mohit}\ \bibnamefont
  {Randeria}},\ }\bibfield  {title} {\enquote {\bibinfo {title}
  {Superconductor-insulator transition in a disordered electronic system},}\
  }\href {\doibase 10.1103/PhysRevB.54.R3756} {\bibfield  {journal} {\bibinfo
  {journal} {Phys. Rev. B}\ }\textbf {\bibinfo {volume} {54}},\ \bibinfo
  {pages} {R3756--R3759} (\bibinfo {year} {1996})}\BibitemShut {NoStop}%
\bibitem [{\citenamefont {Chen}\ \emph {et~al.}(2018)\citenamefont {Chen},
  \citenamefont {Xu}, \citenamefont {Liu}, \citenamefont {Batrouni},
  \citenamefont {Scalettar},\ and\ \citenamefont {Meng}}]{Chen18}%
  \BibitemOpen
  \bibfield  {author} {\bibinfo {author} {\bibfnamefont {C.}~\bibnamefont
  {Chen}}, \bibinfo {author} {\bibfnamefont {X.Y.}\ \bibnamefont {Xu}},
  \bibinfo {author} {\bibfnamefont {J.}~\bibnamefont {Liu}}, \bibinfo {author}
  {\bibfnamefont {G.}~\bibnamefont {Batrouni}}, \bibinfo {author}
  {\bibfnamefont {R.}~\bibnamefont {Scalettar}}, \ and\ \bibinfo {author}
  {\bibfnamefont {Z.Y.}\ \bibnamefont {Meng}},\ }\bibfield  {title} {\enquote
  {\bibinfo {title} {{Symmetry-enforced self-learning Monte Carlo method
  applied to the Holstein model}},}\ }\href {\doibase
  10.1103/PhysRevB.98.041102} {\bibfield  {journal} {\bibinfo  {journal} {Phys.
  Rev. B}\ }\textbf {\bibinfo {volume} {98}},\ \bibinfo {pages} {041102}
  (\bibinfo {year} {2018})}\BibitemShut {NoStop}%
\bibitem [{\citenamefont {Li}\ \emph {et~al.}(2019)\citenamefont {Li},
  \citenamefont {Dee}, \citenamefont {Khatami},\ and\ \citenamefont
  {Johnston}}]{Shaozhi19}%
  \BibitemOpen
  \bibfield  {author} {\bibinfo {author} {\bibfnamefont {S.}~\bibnamefont
  {Li}}, \bibinfo {author} {\bibfnamefont {P.M.}\ \bibnamefont {Dee}}, \bibinfo
  {author} {\bibfnamefont {E.}~\bibnamefont {Khatami}}, \ and\ \bibinfo
  {author} {\bibfnamefont {S.}~\bibnamefont {Johnston}},\ }\bibfield  {title}
  {\enquote {\bibinfo {title} {{Accelerating lattice quantum Monte Carlo
  simulations using artificial neural networks: Application to the Holstein
  model}},}\ }\href {\doibase 10.1103/PhysRevB.100.020302} {\bibfield
  {journal} {\bibinfo  {journal} {Phys. Rev. B}\ }\textbf {\bibinfo {volume}
  {100}},\ \bibinfo {pages} {020302} (\bibinfo {year} {2019})}\BibitemShut
  {NoStop}%
\bibitem [{\citenamefont {Berger}\ \emph {et~al.}(1995)\citenamefont {Berger},
  \citenamefont {Val\'a\ifmmode~\check{s}\else \v{s}\fi{}ek},\ and\
  \citenamefont {von~der Linden}}]{Berger95}%
  \BibitemOpen
  \bibfield  {author} {\bibinfo {author} {\bibfnamefont {E.}~\bibnamefont
  {Berger}}, \bibinfo {author} {\bibfnamefont {P.}~\bibnamefont
  {Val\'a\ifmmode~\check{s}\else \v{s}\fi{}ek}}, \ and\ \bibinfo {author}
  {\bibfnamefont {W.}~\bibnamefont {von~der Linden}},\ }\bibfield  {title}
  {\enquote {\bibinfo {title} {{Two-dimensional Hubbard-Holstein model}},}\
  }\href {\doibase 10.1103/PhysRevB.52.4806} {\bibfield  {journal} {\bibinfo
  {journal} {Phys. Rev. B}\ }\textbf {\bibinfo {volume} {52}},\ \bibinfo
  {pages} {4806--4814} (\bibinfo {year} {1995})}\BibitemShut {NoStop}%
\bibitem [{\citenamefont {Varney}\ \emph {et~al.}(2009)\citenamefont {Varney},
  \citenamefont {Lee}, \citenamefont {Bai}, \citenamefont {Chiesa},
  \citenamefont {Jarrell},\ and\ \citenamefont {Scalettar}}]{varney09}%
  \BibitemOpen
  \bibfield  {author} {\bibinfo {author} {\bibfnamefont {C.N.}\ \bibnamefont
  {Varney}}, \bibinfo {author} {\bibfnamefont {C.R.}\ \bibnamefont {Lee}},
  \bibinfo {author} {\bibfnamefont {Z.J.}\ \bibnamefont {Bai}}, \bibinfo
  {author} {\bibfnamefont {S.}~\bibnamefont {Chiesa}}, \bibinfo {author}
  {\bibfnamefont {M.}~\bibnamefont {Jarrell}}, \ and\ \bibinfo {author}
  {\bibfnamefont {R.T.}\ \bibnamefont {Scalettar}},\ }\bibfield  {title}
  {\enquote {\bibinfo {title} {{Quantum Monte Carlo Study of the 2D Fermion
  Hubbard Model at Half-Filling}},}\ }\href@noop {} {\bibfield  {journal}
  {\bibinfo  {journal} {Phys. Rev. B}\ }\textbf {\bibinfo {volume} {80}},\
  \bibinfo {pages} {075116} (\bibinfo {year} {2009})}\BibitemShut {NoStop}%
\bibitem [{\citenamefont {Scalettar}\ \emph
  {et~al.}(1989{\natexlab{b}})\citenamefont {Scalettar}, \citenamefont {Loh},
  \citenamefont {Gubernatis}, \citenamefont {Moreo}, \citenamefont {White},
  \citenamefont {Scalapino}, \citenamefont {Sugar},\ and\ \citenamefont
  {Dagotto}}]{scalettar89b}%
  \BibitemOpen
  \bibfield  {author} {\bibinfo {author} {\bibfnamefont {R.T.}\ \bibnamefont
  {Scalettar}}, \bibinfo {author} {\bibfnamefont {E.Y.}\ \bibnamefont {Loh}},
  \bibinfo {author} {\bibfnamefont {J.E.}\ \bibnamefont {Gubernatis}}, \bibinfo
  {author} {\bibfnamefont {A.}~\bibnamefont {Moreo}}, \bibinfo {author}
  {\bibfnamefont {S.R.}\ \bibnamefont {White}}, \bibinfo {author}
  {\bibfnamefont {D.J.}\ \bibnamefont {Scalapino}}, \bibinfo {author}
  {\bibfnamefont {R.L.}\ \bibnamefont {Sugar}}, \ and\ \bibinfo {author}
  {\bibfnamefont {E.}~\bibnamefont {Dagotto}},\ }\bibfield  {title} {\enquote
  {\bibinfo {title} {{Phase Diagram of the Two-Dimensional Negative U Hubbard
  Model}},}\ }\href@noop {} {\bibfield  {journal} {\bibinfo  {journal} {Phys.
  Rev. Lett.}\ }\textbf {\bibinfo {volume} {62}},\ \bibinfo {pages} {1407}
  (\bibinfo {year} {1989}{\natexlab{b}})}\BibitemShut {NoStop}%
\bibitem [{\citenamefont {Bouadim}\ \emph {et~al.}(2011)\citenamefont
  {Bouadim}, \citenamefont {Loh}, \citenamefont {Randeria},\ and\ \citenamefont
  {Trivedi}}]{bouadim11}%
  \BibitemOpen
  \bibfield  {author} {\bibinfo {author} {\bibfnamefont {K.}~\bibnamefont
  {Bouadim}}, \bibinfo {author} {\bibfnamefont {Y.~L.}\ \bibnamefont {Loh}},
  \bibinfo {author} {\bibfnamefont {M.}~\bibnamefont {Randeria}}, \ and\
  \bibinfo {author} {\bibfnamefont {N.}~\bibnamefont {Trivedi}},\ }\bibfield
  {title} {\enquote {\bibinfo {title} {Single- and two-particle energy gaps
  across the disorder-driven superconductor-insulator transition},}\
  }\href@noop {} {\bibfield  {journal} {\bibinfo  {journal} {Nature Physics}\
  }\textbf {\bibinfo {volume} {7}},\ \bibinfo {pages} {884} (\bibinfo {year}
  {2011})}\BibitemShut {NoStop}%
\bibitem [{\citenamefont {Esterlis}\ \emph
  {et~al.}(2018{\natexlab{a}})\citenamefont {Esterlis}, \citenamefont
  {Kivelson},\ and\ \citenamefont {Scalapino}}]{esterlis18b}%
  \BibitemOpen
  \bibfield  {author} {\bibinfo {author} {\bibfnamefont {I.}~\bibnamefont
  {Esterlis}}, \bibinfo {author} {\bibfnamefont {S.}~\bibnamefont {Kivelson}},
  \ and\ \bibinfo {author} {\bibfnamefont {D.}~\bibnamefont {Scalapino}},\
  }\bibfield  {title} {\enquote {\bibinfo {title} {A bound on the
  superconducting transition temperature},}\ }\href@noop {} {\bibfield
  {journal} {\bibinfo  {journal} {Nature Physics Journal, Quantum Materials}\
  }\textbf {\bibinfo {volume} {3}},\ \bibinfo {pages} {59} (\bibinfo {year}
  {2018}{\natexlab{a}})}\BibitemShut {NoStop}%
\bibitem [{\citenamefont {Veki\ifmmode~\acute{c}\else \'{c}\fi{}}\ \emph
  {et~al.}(1992)\citenamefont {Veki\ifmmode~\acute{c}\else \'{c}\fi{}},
  \citenamefont {Noack},\ and\ \citenamefont {White}}]{Vekic92}%
  \BibitemOpen
  \bibfield  {author} {\bibinfo {author} {\bibfnamefont {M.}~\bibnamefont
  {Veki\ifmmode~\acute{c}\else \'{c}\fi{}}}, \bibinfo {author} {\bibfnamefont
  {R.~M.}\ \bibnamefont {Noack}}, \ and\ \bibinfo {author} {\bibfnamefont
  {S.~R.}\ \bibnamefont {White}},\ }\bibfield  {title} {\enquote {\bibinfo
  {title} {{Charge-density waves versus superconductivity in the Holstein model
  with next-nearest-neighbor hopping}},}\ }\href {\doibase
  10.1103/PhysRevB.46.271} {\bibfield  {journal} {\bibinfo  {journal} {Phys.
  Rev. B}\ }\textbf {\bibinfo {volume} {46}},\ \bibinfo {pages} {271--278}
  (\bibinfo {year} {1992})}\BibitemShut {NoStop}%
\bibitem [{\citenamefont {Freericks}\ \emph {et~al.}(1993)\citenamefont
  {Freericks}, \citenamefont {Jarrell},\ and\ \citenamefont
  {Scalapino}}]{Freericks93}%
  \BibitemOpen
  \bibfield  {author} {\bibinfo {author} {\bibfnamefont {J.~K.}\ \bibnamefont
  {Freericks}}, \bibinfo {author} {\bibfnamefont {M.}~\bibnamefont {Jarrell}},
  \ and\ \bibinfo {author} {\bibfnamefont {D.~J.}\ \bibnamefont {Scalapino}},\
  }\bibfield  {title} {\enquote {\bibinfo {title} {Holstein model in infinite
  dimensions},}\ }\href {\doibase 10.1103/PhysRevB.48.6302} {\bibfield
  {journal} {\bibinfo  {journal} {Phys. Rev. B}\ }\textbf {\bibinfo {volume}
  {48}},\ \bibinfo {pages} {6302--6314} (\bibinfo {year} {1993})}\BibitemShut
  {NoStop}%
\bibitem [{\citenamefont {Costa}\ \emph {et~al.}(2018)\citenamefont {Costa},
  \citenamefont {Blommel}, \citenamefont {Chiu}, \citenamefont {Batrouni},\
  and\ \citenamefont {Scalettar}}]{Costa18}%
  \BibitemOpen
  \bibfield  {author} {\bibinfo {author} {\bibfnamefont {N.~C.}\ \bibnamefont
  {Costa}}, \bibinfo {author} {\bibfnamefont {T.}~\bibnamefont {Blommel}},
  \bibinfo {author} {\bibfnamefont {W.-T.}\ \bibnamefont {Chiu}}, \bibinfo
  {author} {\bibfnamefont {G.}~\bibnamefont {Batrouni}}, \ and\ \bibinfo
  {author} {\bibfnamefont {R.~T.}\ \bibnamefont {Scalettar}},\ }\bibfield
  {title} {\enquote {\bibinfo {title} {Phonon dispersion and the competition
  between pairing and charge order},}\ }\href {\doibase
  10.1103/PhysRevLett.120.187003} {\bibfield  {journal} {\bibinfo  {journal}
  {Phys. Rev. Lett.}\ }\textbf {\bibinfo {volume} {120}},\ \bibinfo {pages}
  {187003} (\bibinfo {year} {2018})}\BibitemShut {NoStop}%
\bibitem [{\citenamefont {Esterlis}\ \emph
  {et~al.}(2018{\natexlab{b}})\citenamefont {Esterlis}, \citenamefont
  {Nosarzewski}, \citenamefont {Huang}, \citenamefont {Moritz}, \citenamefont
  {Devereaux}, \citenamefont {Scalapino},\ and\ \citenamefont
  {Kivelson}}]{esterlis18}%
  \BibitemOpen
  \bibfield  {author} {\bibinfo {author} {\bibfnamefont {I.}~\bibnamefont
  {Esterlis}}, \bibinfo {author} {\bibfnamefont {B.}~\bibnamefont
  {Nosarzewski}}, \bibinfo {author} {\bibfnamefont {E.~W.}\ \bibnamefont
  {Huang}}, \bibinfo {author} {\bibfnamefont {B.}~\bibnamefont {Moritz}},
  \bibinfo {author} {\bibfnamefont {T.~P.}\ \bibnamefont {Devereaux}}, \bibinfo
  {author} {\bibfnamefont {D.~J.}\ \bibnamefont {Scalapino}}, \ and\ \bibinfo
  {author} {\bibfnamefont {S.~A.}\ \bibnamefont {Kivelson}},\ }\bibfield
  {title} {\enquote {\bibinfo {title} {{Breakdown of the Migdal-Eliashberg
  theory: A determinant quantum Monte Carlo study}},}\ }\href {\doibase
  10.1103/PhysRevB.97.140501} {\bibfield  {journal} {\bibinfo  {journal} {Phys.
  Rev. B}\ }\textbf {\bibinfo {volume} {97}},\ \bibinfo {pages} {140501}
  (\bibinfo {year} {2018}{\natexlab{b}})}\BibitemShut {NoStop}%
\bibitem [{\citenamefont {Yamazaki}\ \emph {et~al.}(2014)\citenamefont
  {Yamazaki}, \citenamefont {Hoshino},\ and\ \citenamefont
  {Kuramoto}}]{yamazaki14}%
  \BibitemOpen
  \bibfield  {author} {\bibinfo {author} {\bibfnamefont {S.}~\bibnamefont
  {Yamazaki}}, \bibinfo {author} {\bibfnamefont {S.}~\bibnamefont {Hoshino}}, \
  and\ \bibinfo {author} {\bibfnamefont {Y.}~\bibnamefont {Kuramoto}},\
  }\bibfield  {title} {\enquote {\bibinfo {title} {{Continuous-Time Quantum
  Monte Carlo Study of Strong Coupling Superconductivity in Holstein-Hubbard
  Model}},}\ }\href {\doibase 10.7566/JPSCP.3.016021} {\bibfield  {journal}
  {\bibinfo  {journal} {JPS Conf. Proc.}\ }\textbf {\bibinfo {volume} {3}},\
  \bibinfo {pages} {016021} (\bibinfo {year} {2014})}\BibitemShut {NoStop}%
\bibitem [{\citenamefont {Karakuzu}\ \emph {et~al.}(2017)\citenamefont
  {Karakuzu}, \citenamefont {Tocchio}, \citenamefont {Sorella},\ and\
  \citenamefont {Becca}}]{Karakuzu17}%
  \BibitemOpen
  \bibfield  {author} {\bibinfo {author} {\bibfnamefont {S.}~\bibnamefont
  {Karakuzu}}, \bibinfo {author} {\bibfnamefont {L.F.}\ \bibnamefont
  {Tocchio}}, \bibinfo {author} {\bibfnamefont {S.}~\bibnamefont {Sorella}}, \
  and\ \bibinfo {author} {\bibfnamefont {F.}~\bibnamefont {Becca}},\ }\bibfield
   {title} {\enquote {\bibinfo {title} {{Superconductivity, charge-density
  waves, antiferromagnetism, and phase separation in the Hubbard-Holstein
  model}},}\ }\href {\doibase 10.1103/PhysRevB.96.205145} {\bibfield  {journal}
  {\bibinfo  {journal} {Phys. Rev. B}\ }\textbf {\bibinfo {volume} {96}},\
  \bibinfo {pages} {205145} (\bibinfo {year} {2017})}\BibitemShut {NoStop}%
\bibitem [{\citenamefont {Ohgoe}\ and\ \citenamefont {Imada}(2017)}]{Ohgoe17}%
  \BibitemOpen
  \bibfield  {author} {\bibinfo {author} {\bibfnamefont {T.}~\bibnamefont
  {Ohgoe}}\ and\ \bibinfo {author} {\bibfnamefont {M.}~\bibnamefont {Imada}},\
  }\bibfield  {title} {\enquote {\bibinfo {title} {{Competition among
  superconducting, antiferromagnetic, and charge orders with intervention by
  phase separation in the 2D Holstein-Hubbard model}},}\ }\href {\doibase
  10.1103/PhysRevLett.119.197001} {\bibfield  {journal} {\bibinfo  {journal}
  {Phys. Rev. Lett.}\ }\textbf {\bibinfo {volume} {119}},\ \bibinfo {pages}
  {197001} (\bibinfo {year} {2017})}\BibitemShut {NoStop}%
\bibitem [{\citenamefont {Costa}\ \emph
  {et~al.}(2020{\natexlab{a}})\citenamefont {Costa}, \citenamefont {Seki},
  \citenamefont {Yunoki},\ and\ \citenamefont {Sorella}}]{Costa20}%
  \BibitemOpen
  \bibfield  {author} {\bibinfo {author} {\bibfnamefont {Natanael~C}\
  \bibnamefont {Costa}}, \bibinfo {author} {\bibfnamefont {Kazuhiro}\
  \bibnamefont {Seki}}, \bibinfo {author} {\bibfnamefont {Seiji}\ \bibnamefont
  {Yunoki}}, \ and\ \bibinfo {author} {\bibfnamefont {Sandro}\ \bibnamefont
  {Sorella}},\ }\bibfield  {title} {\enquote {\bibinfo {title} {{Phase diagram
  of the two-dimensional Hubbard-Holstein model}},}\ }\href {\doibase
  10.1038/s42005-020-0342-2} {\bibfield  {journal} {\bibinfo  {journal}
  {Communications Physics}\ }\textbf {\bibinfo {volume} {3}},\ \bibinfo {pages}
  {1--6} (\bibinfo {year} {2020}{\natexlab{a}})}\BibitemShut {NoStop}%
\bibitem [{\citenamefont {Wang}\ \emph {et~al.}(2020)\citenamefont {Wang},
  \citenamefont {Esterlis}, \citenamefont {Shi}, \citenamefont {Cirac},\ and\
  \citenamefont {Demler}}]{Wang19}%
  \BibitemOpen
  \bibfield  {author} {\bibinfo {author} {\bibfnamefont {Yao}\ \bibnamefont
  {Wang}}, \bibinfo {author} {\bibfnamefont {Ilya}\ \bibnamefont {Esterlis}},
  \bibinfo {author} {\bibfnamefont {Tao}\ \bibnamefont {Shi}}, \bibinfo
  {author} {\bibfnamefont {J.~Ignacio}\ \bibnamefont {Cirac}}, \ and\ \bibinfo
  {author} {\bibfnamefont {Eugene}\ \bibnamefont {Demler}},\ }\bibfield
  {title} {\enquote {\bibinfo {title} {Zero-temperature phases of the
  two-dimensional hubbard-holstein model: A non-gaussian exact diagonalization
  study},}\ }\href {\doibase 10.1103/PhysRevResearch.2.043258} {\bibfield
  {journal} {\bibinfo  {journal} {Phys. Rev. Research}\ }\textbf {\bibinfo
  {volume} {2}},\ \bibinfo {pages} {043258} (\bibinfo {year}
  {2020})}\BibitemShut {NoStop}%
\bibitem [{\citenamefont {Costa}\ \emph
  {et~al.}(2020{\natexlab{b}})\citenamefont {Costa}, \citenamefont {Seki},\
  and\ \citenamefont {Sorella}}]{Costa20b}%
  \BibitemOpen
  \bibfield  {author} {\bibinfo {author} {\bibfnamefont {Natanael~C.}\
  \bibnamefont {Costa}}, \bibinfo {author} {\bibfnamefont {Kazuhiro}\
  \bibnamefont {Seki}}, \ and\ \bibinfo {author} {\bibfnamefont {Sandro}\
  \bibnamefont {Sorella}},\ }\bibfield  {title} {\enquote {\bibinfo {title}
  {Magnetism and charge order in the honeycomb lattice},}\ }\href@noop {}
  {\bibfield  {journal} {\bibinfo  {journal} {arXiv:2009.05586}\ } (\bibinfo
  {year} {2020}{\natexlab{b}})}\BibitemShut {NoStop}%
\bibitem [{\citenamefont {Bronold}\ and\ \citenamefont
  {Fehske}(2002)}]{Bronold02}%
  \BibitemOpen
  \bibfield  {author} {\bibinfo {author} {\bibfnamefont {F.X.}\ \bibnamefont
  {Bronold}}\ and\ \bibinfo {author} {\bibfnamefont {H.}~\bibnamefont
  {Fehske}},\ }\bibfield  {title} {\enquote {\bibinfo {title} {Anderson
  localization of polaron states},}\ }\href {\doibase
  10.1103/PhysRevB.66.073102} {\bibfield  {journal} {\bibinfo  {journal} {Phys.
  Rev. B}\ }\textbf {\bibinfo {volume} {66}},\ \bibinfo {pages} {073102}
  (\bibinfo {year} {2002})}\BibitemShut {NoStop}%
\bibitem [{\citenamefont {Ebrahimnejad}\ and\ \citenamefont
  {Berciu}(2012{\natexlab{a}})}]{Ebrahimnejad12}%
  \BibitemOpen
  \bibfield  {author} {\bibinfo {author} {\bibfnamefont {H.}~\bibnamefont
  {Ebrahimnejad}}\ and\ \bibinfo {author} {\bibfnamefont {M.}~\bibnamefont
  {Berciu}},\ }\bibfield  {title} {\enquote {\bibinfo {title} {{Perturbational
  study of the lifetime of a Holstein polaron in the presence of weak
  disorder}},}\ }\href {\doibase 10.1103/PhysRevB.86.205109} {\bibfield
  {journal} {\bibinfo  {journal} {Phys. Rev. B}\ }\textbf {\bibinfo {volume}
  {86}},\ \bibinfo {pages} {205109} (\bibinfo {year}
  {2012}{\natexlab{a}})}\BibitemShut {NoStop}%
\bibitem [{\citenamefont {Ebrahimnejad}\ and\ \citenamefont
  {Berciu}(2012{\natexlab{b}})}]{Ebrahimnejad12b}%
  \BibitemOpen
  \bibfield  {author} {\bibinfo {author} {\bibfnamefont {H.}~\bibnamefont
  {Ebrahimnejad}}\ and\ \bibinfo {author} {\bibfnamefont {M.}~\bibnamefont
  {Berciu}},\ }\bibfield  {title} {\enquote {\bibinfo {title} {{Trapping of
  three-dimensional Holstein polarons by various impurities}},}\ }\href
  {\doibase 10.1103/PhysRevB.85.165117} {\bibfield  {journal} {\bibinfo
  {journal} {Phys. Rev. B}\ }\textbf {\bibinfo {volume} {85}},\ \bibinfo
  {pages} {165117} (\bibinfo {year} {2012}{\natexlab{b}})}\BibitemShut
  {NoStop}%
\bibitem [{\citenamefont {Tozer}\ and\ \citenamefont
  {Barford}(2014)}]{Tozer14}%
  \BibitemOpen
  \bibfield  {author} {\bibinfo {author} {\bibfnamefont {O.R.}\ \bibnamefont
  {Tozer}}\ and\ \bibinfo {author} {\bibfnamefont {W.}~\bibnamefont
  {Barford}},\ }\bibfield  {title} {\enquote {\bibinfo {title} {{Localization
  of large polarons in the disordered Holstein model}},}\ }\href {\doibase
  10.1103/PhysRevB.89.155434} {\bibfield  {journal} {\bibinfo  {journal} {Phys.
  Rev. B}\ }\textbf {\bibinfo {volume} {89}},\ \bibinfo {pages} {155434}
  (\bibinfo {year} {2014})}\BibitemShut {NoStop}%
\bibitem [{\citenamefont {Zhang}\ \emph {et~al.}()\citenamefont {Zhang},
  \citenamefont {Feng}, \citenamefont {Batrouni},\ and\ \citenamefont
  {Scalettar}}]{zhang20}%
  \BibitemOpen
  \bibfield  {author} {\bibinfo {author} {\bibfnamefont {Y.}~\bibnamefont
  {Zhang}}, \bibinfo {author} {\bibfnamefont {C.}~\bibnamefont {Feng}},
  \bibinfo {author} {\bibfnamefont {G.}~\bibnamefont {Batrouni}}, \ and\
  \bibinfo {author} {\bibfnamefont {R.}~\bibnamefont {Scalettar}},\ }\href@noop
  {} {\bibinfo  {journal} {work in progress}\ }\BibitemShut {NoStop}%
\bibitem [{\citenamefont {Hohenadler}\ and\ \citenamefont
  {Batrouni}(2019)}]{hohenadler19}%
  \BibitemOpen
\bibfield  {journal} {  }\bibfield  {author} {\bibinfo {author} {\bibfnamefont
  {M.}~\bibnamefont {Hohenadler}}\ and\ \bibinfo {author} {\bibfnamefont
  {G.~G.}\ \bibnamefont {Batrouni}},\ }\bibfield  {title} {\enquote {\bibinfo
  {title} {{Dominant charge-density-wave correlations in the Holstein model on
  the half-filled square lattice}},}\ }\href@noop {} {\bibfield  {journal}
  {\bibinfo  {journal} {Phys. Rev.}\ }\textbf {\bibinfo {volume} {B100}},\
  \bibinfo {pages} {165114} (\bibinfo {year} {2019})}\BibitemShut {NoStop}%
\bibitem [{\citenamefont {McMahan}\ \emph {et~al.}(1998)\citenamefont
  {McMahan}, \citenamefont {Huscroft}, \citenamefont {Scalettar},\ and\
  \citenamefont {Pollock}}]{mcmahan98}%
  \BibitemOpen
  \bibfield  {author} {\bibinfo {author} {\bibfnamefont {A.}~\bibnamefont
  {McMahan}}, \bibinfo {author} {\bibfnamefont {C.}~\bibnamefont {Huscroft}},
  \bibinfo {author} {\bibfnamefont {R.T.}\ \bibnamefont {Scalettar}}, \ and\
  \bibinfo {author} {\bibfnamefont {E.L.}\ \bibnamefont {Pollock}},\ }\bibfield
   {title} {\enquote {\bibinfo {title} {{Volume Collapse transitions in the
  rare earth metals}},}\ }\href@noop {} {\bibfield  {journal} {\bibinfo
  {journal} {J. of Computer-Aided Materials Design}\ }\textbf {\bibinfo
  {volume} {5}},\ \bibinfo {pages} {131} (\bibinfo {year} {1998})}\BibitemShut
  {NoStop}%
\bibitem [{\citenamefont {Paiva}\ \emph {et~al.}(2001)\citenamefont {Paiva},
  \citenamefont {Scalettar}, \citenamefont {Huscroft},\ and\ \citenamefont
  {McMahan}}]{paiva01}%
  \BibitemOpen
  \bibfield  {author} {\bibinfo {author} {\bibfnamefont {T.}~\bibnamefont
  {Paiva}}, \bibinfo {author} {\bibfnamefont {R.~T.}\ \bibnamefont
  {Scalettar}}, \bibinfo {author} {\bibfnamefont {C.}~\bibnamefont {Huscroft}},
  \ and\ \bibinfo {author} {\bibfnamefont {A.~K.}\ \bibnamefont {McMahan}},\
  }\bibfield  {title} {\enquote {\bibinfo {title} {{Signatures of spin and
  charge energy scales in the local moment and specific heat of the half-filled
  two-dimensional Hubbard model}},}\ }\href {\doibase
  10.1103/PhysRevB.63.125116} {\bibfield  {journal} {\bibinfo  {journal} {Phys.
  Rev. B}\ }\textbf {\bibinfo {volume} {63}},\ \bibinfo {pages} {125116}
  (\bibinfo {year} {2001})}\BibitemShut {NoStop}%
\bibitem [{\citenamefont {Ulmke}\ and\ \citenamefont
  {Scalettar}(1997)}]{ulmke97}%
  \BibitemOpen
  \bibfield  {author} {\bibinfo {author} {\bibfnamefont {M.}~\bibnamefont
  {Ulmke}}\ and\ \bibinfo {author} {\bibfnamefont {R.}~\bibnamefont
  {Scalettar}},\ }\bibfield  {title} {\enquote {\bibinfo {title} {{Magnetic
  Correlations in the Two Dimensional Anderson--Hubbard Model}},}\ }\href@noop
  {} {\bibfield  {journal} {\bibinfo  {journal} {Phys. Rev. B}\ }\textbf
  {\bibinfo {volume} {55}},\ \bibinfo {pages} {4149} (\bibinfo {year}
  {1997})}\BibitemShut {NoStop}%
\bibitem [{\citenamefont {Ulmke}\ \emph {et~al.}(1995)\citenamefont {Ulmke},
  \citenamefont {Jani\ifmmode~\check{s}\else \v{s}\fi{}},\ and\ \citenamefont
  {Vollhardt}}]{ulmke95}%
  \BibitemOpen
  \bibfield  {author} {\bibinfo {author} {\bibfnamefont {M.}~\bibnamefont
  {Ulmke}}, \bibinfo {author} {\bibfnamefont {V.}~\bibnamefont
  {Jani\ifmmode~\check{s}\else \v{s}\fi{}}}, \ and\ \bibinfo {author}
  {\bibfnamefont {D.}~\bibnamefont {Vollhardt}},\ }\bibfield  {title} {\enquote
  {\bibinfo {title} {{Anderson-Hubbard model in infinite dimensions}},}\ }\href
  {\doibase 10.1103/PhysRevB.51.10411} {\bibfield  {journal} {\bibinfo
  {journal} {Phys. Rev. B}\ }\textbf {\bibinfo {volume} {51}},\ \bibinfo
  {pages} {10411--10426} (\bibinfo {year} {1995})}\BibitemShut {NoStop}%
\bibitem [{\citenamefont {Bickers}\ \emph {et~al.}(1987)\citenamefont
  {Bickers}, \citenamefont {Scalapino},\ and\ \citenamefont
  {Scalettar}}]{bickers87}%
  \BibitemOpen
  \bibfield  {author} {\bibinfo {author} {\bibfnamefont {N.E.}\ \bibnamefont
  {Bickers}}, \bibinfo {author} {\bibfnamefont {D.J.}\ \bibnamefont
  {Scalapino}}, \ and\ \bibinfo {author} {\bibfnamefont {R.T.}\ \bibnamefont
  {Scalettar}},\ }\bibfield  {title} {\enquote {\bibinfo {title} {Cdw and sdw
  mediated pairing mechanisms},}\ }\href@noop {} {\bibfield  {journal}
  {\bibinfo  {journal} {Int. J. Mod. Phys. B}\ }\textbf {\bibinfo {volume}
  {1}},\ \bibinfo {pages} {687} (\bibinfo {year} {1987})}\BibitemShut {NoStop}%
\bibitem [{\citenamefont {Silver}\ \emph {et~al.}(1990)\citenamefont {Silver},
  \citenamefont {Sivia},\ and\ \citenamefont {Gubernatis}}]{silver90}%
  \BibitemOpen
  \bibfield  {author} {\bibinfo {author} {\bibfnamefont {R.~N.}\ \bibnamefont
  {Silver}}, \bibinfo {author} {\bibfnamefont {D.~S.}\ \bibnamefont {Sivia}}, \
  and\ \bibinfo {author} {\bibfnamefont {J.~E.}\ \bibnamefont {Gubernatis}},\
  }\bibfield  {title} {\enquote {\bibinfo {title} {Maximum-entropy method for
  analytic continuation of quantum monte carlo data},}\ }\href {\doibase
  10.1103/PhysRevB.41.2380} {\bibfield  {journal} {\bibinfo  {journal} {Phys.
  Rev. B}\ }\textbf {\bibinfo {volume} {41}},\ \bibinfo {pages} {2380--2389}
  (\bibinfo {year} {1990})}\BibitemShut {NoStop}%
\bibitem [{\citenamefont {Gubernatis}\ \emph {et~al.}(1991)\citenamefont
  {Gubernatis}, \citenamefont {Jarrell}, \citenamefont {Silver},\ and\
  \citenamefont {Sivia}}]{gubernatis91}%
  \BibitemOpen
  \bibfield  {author} {\bibinfo {author} {\bibfnamefont {J.~E.}\ \bibnamefont
  {Gubernatis}}, \bibinfo {author} {\bibfnamefont {Mark}\ \bibnamefont
  {Jarrell}}, \bibinfo {author} {\bibfnamefont {R.~N.}\ \bibnamefont {Silver}},
  \ and\ \bibinfo {author} {\bibfnamefont {D.~S.}\ \bibnamefont {Sivia}},\
  }\bibfield  {title} {\enquote {\bibinfo {title} {Quantum monte carlo
  simulations and maximum entropy: Dynamics from imaginary-time data},}\ }\href
  {\doibase 10.1103/PhysRevB.44.6011} {\bibfield  {journal} {\bibinfo
  {journal} {Phys. Rev. B}\ }\textbf {\bibinfo {volume} {44}},\ \bibinfo
  {pages} {6011--6029} (\bibinfo {year} {1991})}\BibitemShut {NoStop}%
\bibitem [{\citenamefont {Sandvik}(1998)}]{sandvik98}%
  \BibitemOpen
  \bibfield  {author} {\bibinfo {author} {\bibfnamefont {Anders~W.}\
  \bibnamefont {Sandvik}},\ }\bibfield  {title} {\enquote {\bibinfo {title}
  {Stochastic method for analytic continuation of quantum monte carlo data},}\
  }\href {\doibase 10.1103/PhysRevB.57.10287} {\bibfield  {journal} {\bibinfo
  {journal} {Phys. Rev. B}\ }\textbf {\bibinfo {volume} {57}},\ \bibinfo
  {pages} {10287--10290} (\bibinfo {year} {1998})}\BibitemShut {NoStop}%
\end{thebibliography}%

\clearpage
\renewcommand{\thefigure}{S\arabic{figure}}
\setcounter{figure}{0}

\noindent
These Supplemental Materials provide additional details concerning 
the CDW-SC competition in the anti-adiabatic limit, 
the Determinant Quantum Monte Carlo method, CDW transition in the clean limit, 
specific heat, temperature dependence in the anti-adiabatic limit, 
and disorder dependence in the adiabatic limit.

\noindent
\begin{figure*}[t]
\begin{center} 
    \includegraphics[width=1.2 \columnwidth]{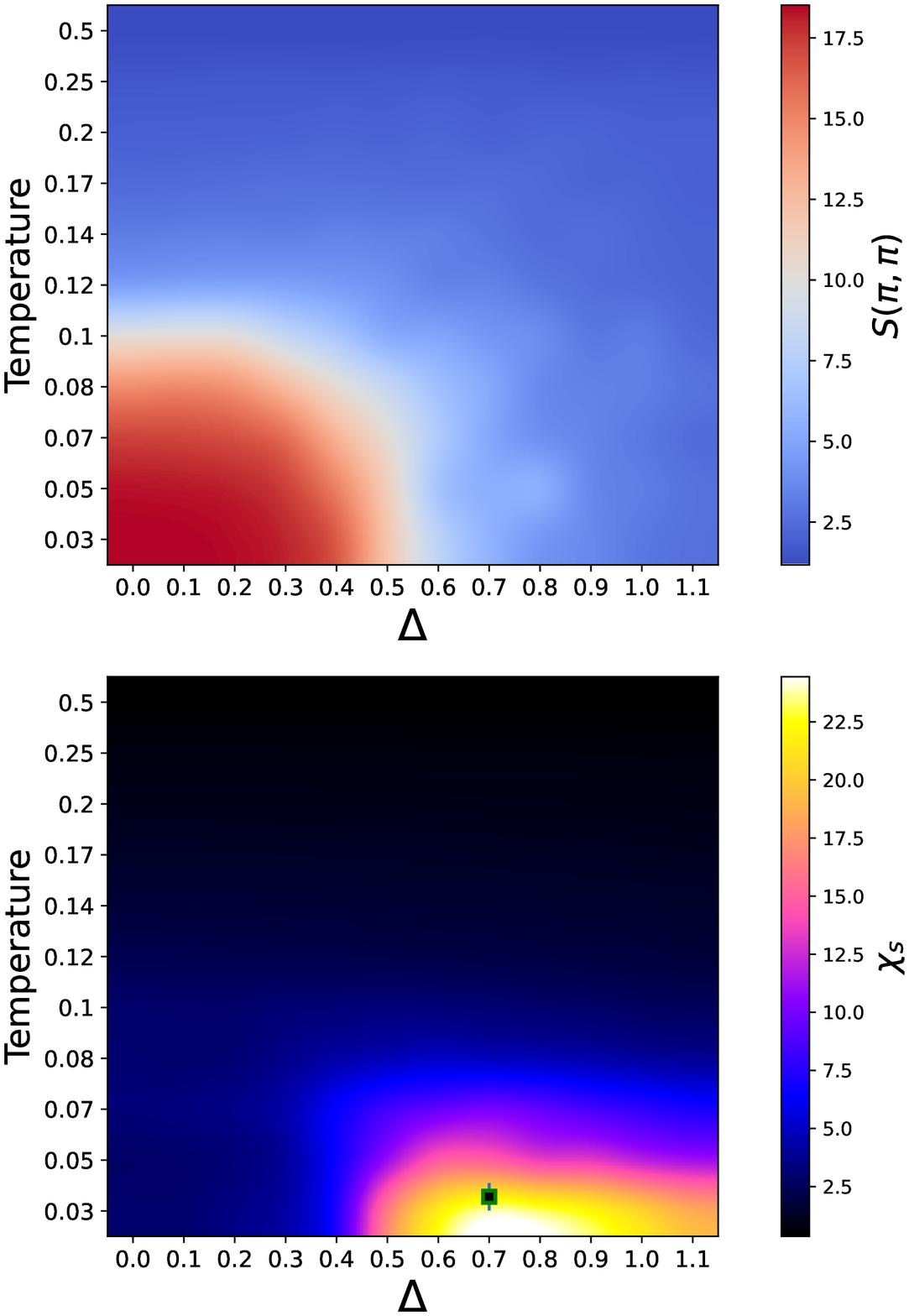} 
    \caption{(a) Heat map of the charge structure factor at ${\bf q}=(\pi,\pi)$, 
      and (b) the pairing susceptibility in the disorder strength-temperature space.  
      Colors correspond to the magnitudes of $S(\pi,\pi)$  and $\chi_{s}$ after interpolation.  
      Here $L=10$, $\omega_{0}=4$ and $\lambda_{D}=0.25$ ($g=2$).  
      To connect this raw heat map data to the onset of superconducting order, 
      we show in the lower panel a symbol representing the transition temperature inferred 
      from finite size scaling of $\chi_{s}$ for different $L$.}
  \label{fig:figS4}
\end{center}
\end{figure*}

\vskip0.10in \noindent
{\bf A: CDW and SC competition in the anti-adiabatic regime-}
In Fig.~\ref{fig:figS4} we plot the heat maps of charge structure
factor $S(\pi,\pi)$ and $\chi_{s}$ for the anti-adiabatic case
$(\omega_{0}=4)$, with weak effective electron-phonon coupling
$(\lambda_{D}=0.25)$.  These show the nature of the dominant
correlations in the disorder strength-temperature plane.  
Therefore, the combination of $S(\pi, \pi)$ 
and $\chi_{s}$ serves as the phase diagram at finite temperature.  
The issue of how the CDW and SC phase meet at temperature below 
$T=0.03$ is beyond the scope of the present set of simulations.  
At low enough temperature, charge order dominates.  
Increasing the strength of
disorder suppresses the CDW.  Instead, SC emerges as the disorder
strength increases at $T < t/20$.  Further increase of disorder
strength ultimately suppresses the SC phase.  These data suggest there
might be a narrow region where CDW and SC exist simultaneously.
However, conclusive evidence for this would require a simultaneous
finite size extrapolation of $S(\pi,\pi)$ and $\chi_{s}$ which
is beyond the capability of the simulations at present.

\vskip0.10in \noindent
{\bf B: Determinant Quantum Monte Carlo-}
The Holstein Hamiltonian is quadratic in the fermion degrees of
freedom.  Hence they can be traced out analytically, leaving an
expression for the partition function which depends on the
space-imaginary time configuration $x_{\bf i}(\tau)$ of the quantum
oscillator degrees of
freedom~\cite{Blankenbecler81,Scalettar89,Noack91,Santos03}.  The
explicit results of the trace operation are determinants, one for each
of the two spin species.  Because the coupling is symmetric in the
spin index, these two determinants are identical.  Their product is a
square, and there is no sign problem in the simulations, for any value
of the parameters in the Hamiltonian, filling, temperature, or lattice
size.  All equal imaginary time observables can be expressed in terms
of elements (or products thereof) of the inverse of the
matrix whose determinant is being sampled.  Hence such measurements
are very inexpensive computationally. Unequal time measurements,
including those of the pair susceptibility and conductivity, require a
separate computation of the un-equal time Greens function, and add
considerably to the simulation time.

\begin{figure}[t]
\includegraphics[scale=0.32]{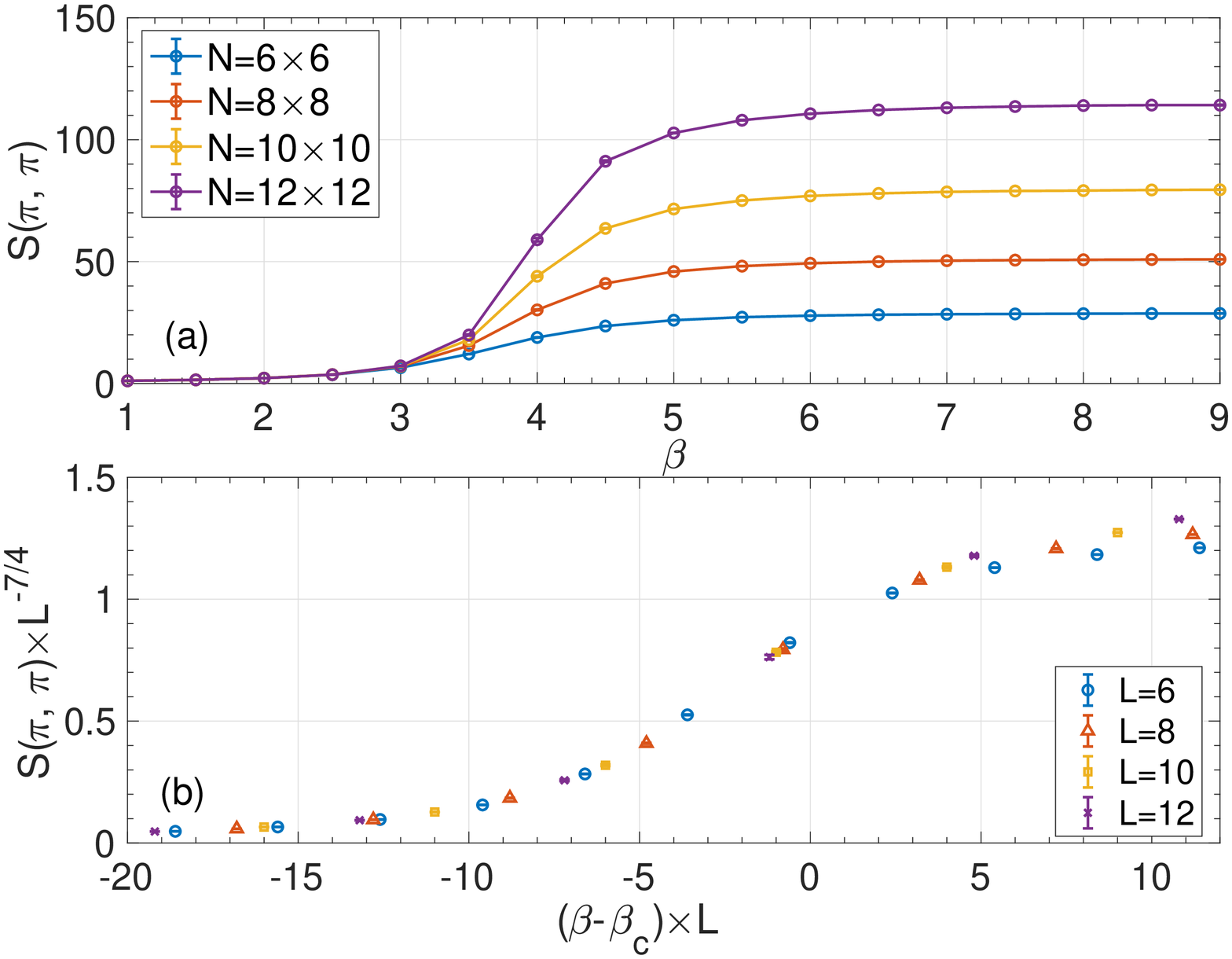}
\caption{(a) The CDW structure factor 
  as a function of inverse temperature $\beta$ for 
  four lattice sizes at $g=1$ and $\omega_0=0.5$
  (b) scaling collapse plot using 2D Ising critical 
  exponents and $\beta_c=4.1$.}
\label{fig:figS1} 
\end{figure}

\vskip0.10in \noindent
{\bf C: CDW transition in the clean limit-}
In the absence of randomness, $\Delta=0$, the half-filled square
lattice Holstein model is believed to undergo a CDW transition for all
values of $\lambda$ and $\omega_0$ as a consequence of the
nesting\cite{hohenadler19} of the Fermi surface and the divergence of
the density of states.  Fig.~\ref{fig:figS1}(a) gives raw data for the
CDW structure factor as a function of inverse temperature $\beta$ for
four lattice sizes at $g=1$ and $\omega_0=0.5$.  At high temperatures
(small $\beta$) the density-density correlation function is short
ranged, only a few local terms contribute to the sum in
Eq.~\ref{eq:Sq} and $S(\pi,\pi)$ is independent of lattice size.  At
low temperatures (large $\beta$) the density correlations extend over
the entire lattice and $S(\pi,\pi)$ grows linearly with volume
$N=L\times L$.  Fig.~\ref{fig:figS1}(b) presents the same data scaled
with the 2D Ising critical exponents, yielding a value for the
transition temperature $T_c \sim 0.24 = 1/ \beta_c$.

\begin{figure}[t]
\includegraphics[scale=0.32]{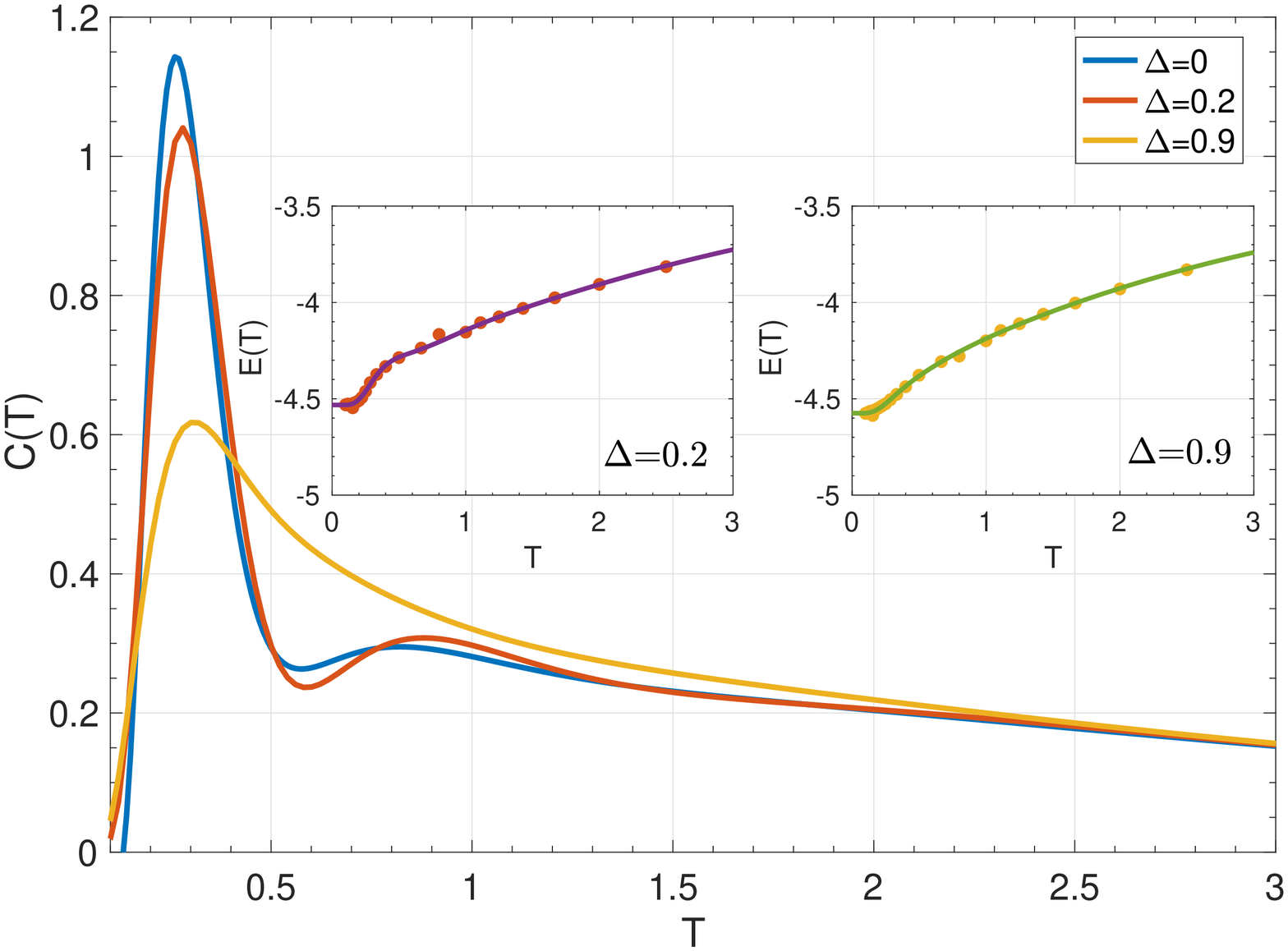}
\caption{Specific heat $C(T)$ as a function of temperature for
  the clean system ($\Delta=0$) and two values of disorder, fixing  
  $L=8$ and $\lambda_{D}=0.5$ ($g=1$).  Inset:
  Raw data for the energy $E(T)$ and the fit given by
  Eq.~\ref{eq:andyfit} at $\Delta=0.2$ and $\Delta=0.9$.}
\label{fig:specific_heat}
\end{figure}

\begin{figure}[t]
\includegraphics[height=2.0in,width=3.2in]{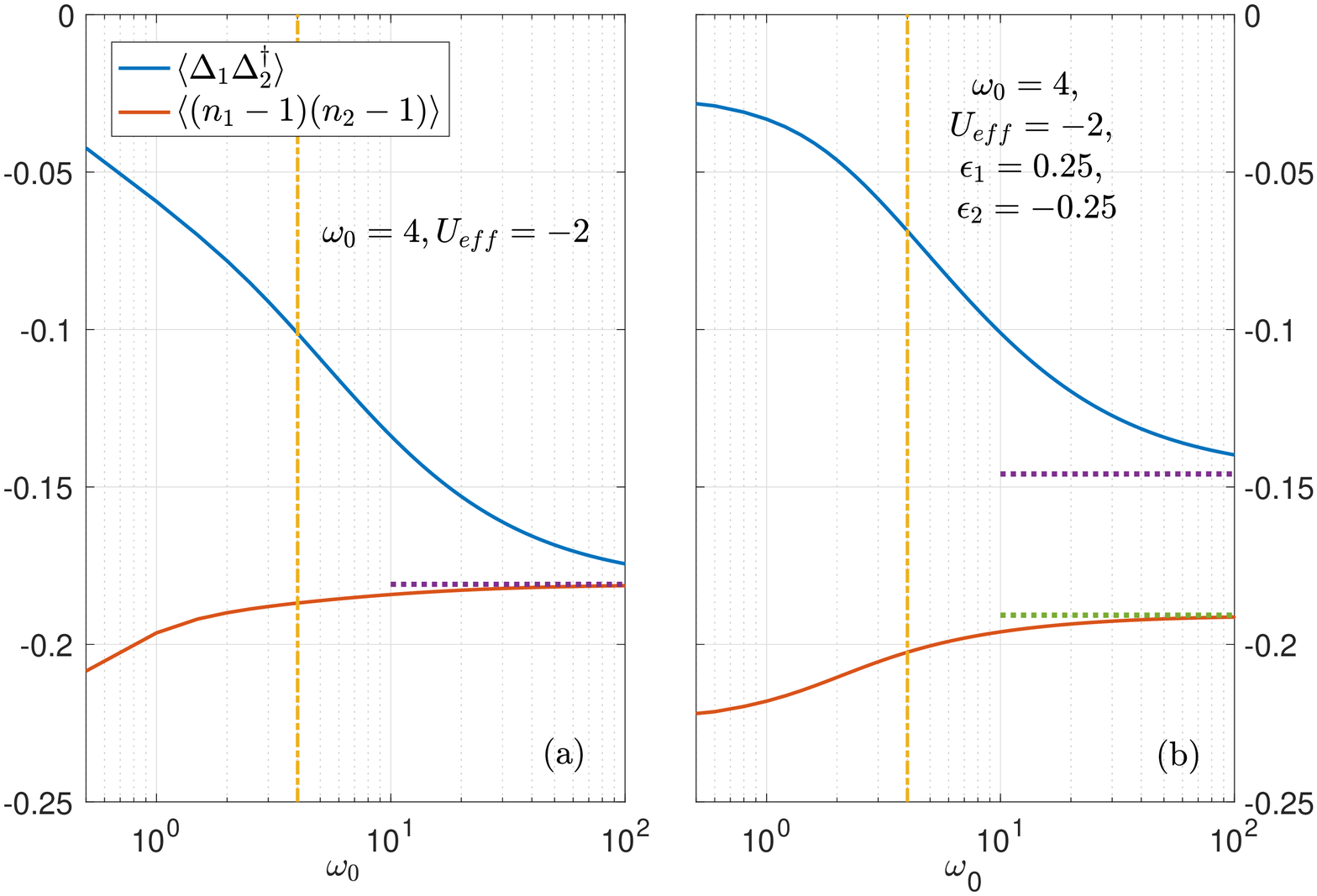}
\caption{
Nearest neighbor pairing and CDW correlations 
for the half-filled Holstein (solid curves) and attractive Hubbard (horizontal
dashed lines at large $\omega_0$) dimers.  
\underbar{Panel a:}  Clean case where the two sites have identical
site energies.  Here the pairing and charge correlations are degenerate
in the Hubbard limit.
\underbar{Panel b:}  `Disordered' case with site energy difference 
$(\epsilon_1-\epsilon_2)/t = 0.50$,
of the same scale as the disorder studied in this paper.
The effect of the site energy is
to break the CDW-Pairing degeneracy (which is already broken 
in the clean Holstein model) also in the Hubbard limit.
In both cases, the Hubbard limit is not reached until
$\omega_0/t \gtrsim 10^2$.  
  } 
\label{fig:holstvshub} 
\end{figure}

\vskip0.10in \noindent
{\bf  D: Relation to Attractive Hubbard Model-}
In light of the known mapping between the Holstein
and Hubbard models in the anti-adiabatic (large $\omega_0$) limit,
it is important to emphasize how our work is distinct
from the previous body of work on the
disordered attractive Hubbard model~\cite{bouadim11}. 
Figure \ref{fig:holstvshub} addresses this issue.  It compares the 
Hubbard and Holstein values for the
nearest neighbor density-density and pair-pair correlations on 
a dimer.  The clean case is shown in panel (a) and with a site energy difference
in panel (b). 
We have chosen an interaction strength $U=-2$ for the attractive Hubbard Hamiltonian, and
vary $g$ and $\omega_0$ together in such a way as to keep
$U_{\rm eff} = -2g^2/\omega_0 =-2$ fixed for the Holstein model.  
While it is true that for $\omega_0/t \rightarrow \infty$
the two models yield the same correlation functions, it is seen that
this limit is only attained for 
$\omega_0/t \gtrsim 10^2$.
Even though the frequencies reported here, $1 < \omega_0/t < 4$
are already high compared to typical phonon frequencies in
real materials, it is clear we are still very far from the Hubbard limit.
Not only are the correlation function values different (by an order of magnitude in the
case of the pairing), but the
CDW-pairing degeneracy of the Hubbard model limit is dramatically broken.
These results demonstrate that the interplay of disorder and interactions presented
here for the Holstein model are expected to
be quite different from the attractive Hubbard model.

Another perspective on the similarities and differences is offered by
considering the interaction between electron mediated by the exchange of 
a phonon propagator,
\begin{equation} \tag{5}
  D(\omega)=\frac{2 g^2 \omega_0}{\omega^2 - \omega_0^2}  \,\,.
\end{equation}
From this expression it is clear that in the limit $\omega_0 >> \omega$ one
recovers an instantaneous attractive interaction whose value matches that of
$U_{\rm eff}$.  Conversely, as one moves away from this anti-adiabatic limit
the electron-phonon interaction will contain frequency dependence not present in
the attractive Hubbard coupling.

In discussing the relation between the two Hamiltonians, it is worth noting that
at low density the Holstein model describes
phenomena such as polaron formation, where a single electron moving on the lattice
has a larger effective mass due to the phonon distortions it carries.  This sort
of physics is not captured by the attractive Hubbard model.  Even though polaron
formation tends to be studied in the dilute limit, the larger effective mass  due
to electron phonon coupling is likely to affect the physics of CDW and SC order at higher
density.  Indeed,
this is one of the reasons the transition temperatures can be quite
different in the Hubbard and Holstein cases (especially at smaller $\omega_0$).

\vskip0.20in \noindent
{\bf E: Specific Heat-}
The effect of disorder on the CDW phase can also be monitored
using thermodynamic responses, most significantly, the specific heat $C(T)$.
To this end, we fit the DQMC
data for the energy per site to the following ansatz~\cite{mcmahan98,paiva01}
\begin{equation} \tag{6}
  E(T)=\omega_0 ~\Big( \frac{1}{e^{\beta \omega_0} -1}
  +\frac{1}{2}\Big)+\sum_{n=1}^{M} c_n e^{-n\beta\delta}~ ,
\label{eq:andyfit}
\end{equation}
in which the parameters $c_n$ and $\delta$ are adjusted to minimize
the deviation of the fitted curve to the data points.  The first term
is the bare energy of the quantum oscillators in the Holstein
Hamiltonian, and the second term captures the electronic
contributions. We then obtain $C(T)$ by differentiating the fitted
expression,  in which we typically set $M=6$ to 8.

Results for the specific heat are shown in Fig.~\ref{fig:specific_heat}.
In the clean limit, $\Delta=0$, $C(T)$ has a broad peak at $T/t \sim
0.8$ corresponding to the temperature scale of pair
formation~\cite{paiva01}, and a sharp peak at $T/t= 0.24 \pm 0.01$ which
aligns well with the critical temperature for the CDW transition
determined by the scaling of $S(\pi,\pi)$ (see Fig.~\ref{fig:figS1} in
the Appendix B).  Similar two-peak structures are observed in the
Hubbard model~\cite{paiva01}, and correspond in that case to the distinct
energy scales of moment formation and antiferromagnetic ordering.  At
weak disorder, $\Delta/t=0.2$, a sharp low temperature peak indicative
of CDW formation persists.  In fact, the peak is first shifted to
slightly higher temperatures.  Such an enhancement of $T_c$ by disorder
has been established in DQMC~\cite{ulmke97} and dynamical mean field
theory~\cite{ulmke95} of the Anderson-Hubbard model.  The effect arises
from the initial growth of the exchange energy $J = 2t^2/(U+\Delta) +
2t^2/(U-\Delta) > 4t^2/U$ with random site energy.  Precisely the same
phenomenon might be expected here, since quantum fluctuations in the CDW
phase have a similar form, with the pair binding energy $4g^2/\omega_0$
playing the role of $U$.  Further increase of $\Delta$ reduces the peak 
of the specific heat, in line with the suppression of the CDW order.

\begin{figure}[t]
\hspace*{-0.7cm}
\includegraphics[scale=0.32]{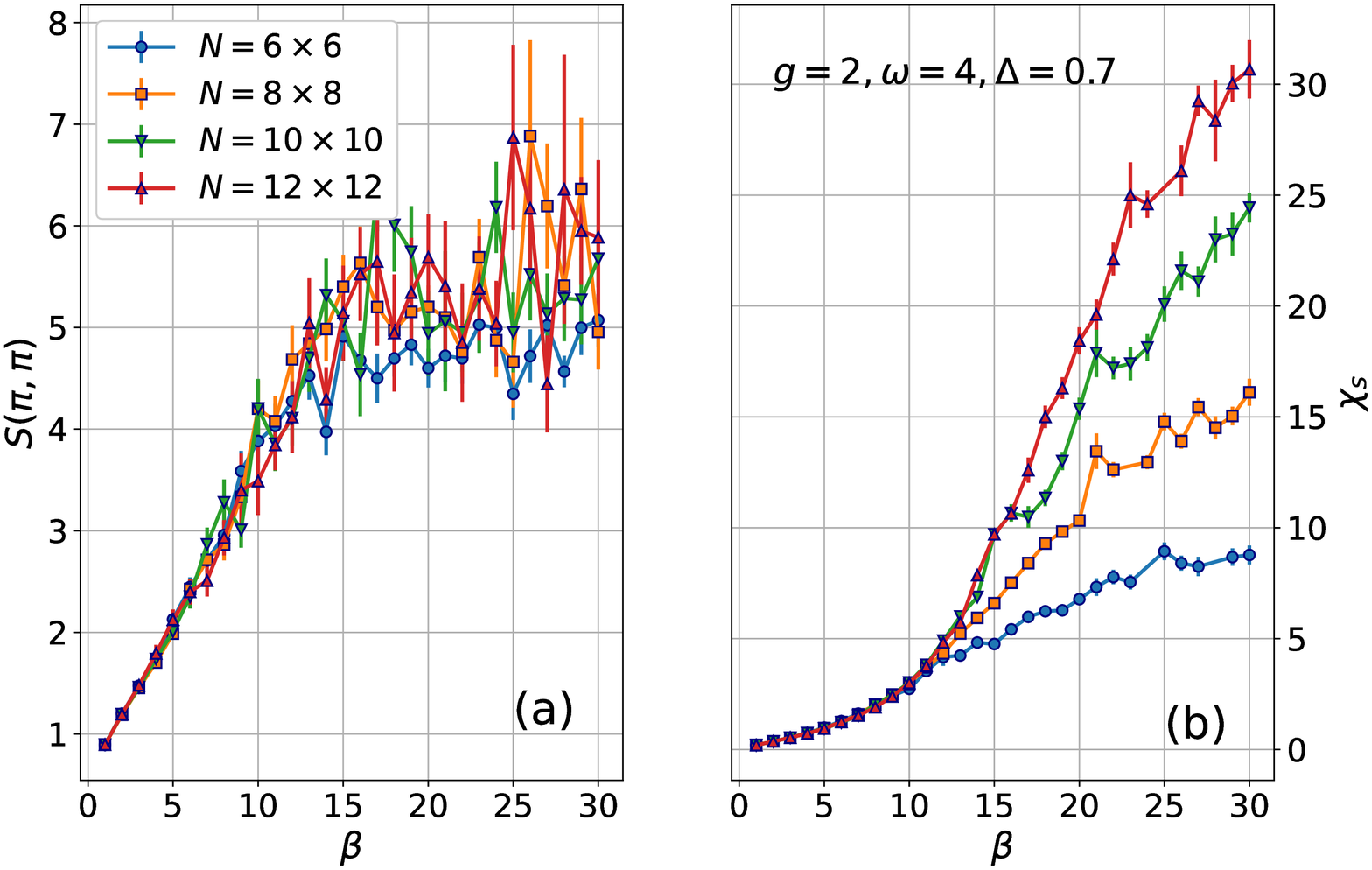}
\caption{The (a) charge structure factor $S(\pi,\pi)$, and (b) s-wave
  pairing susceptibility $\chi_{s}$ as functions of the
  inverse temperature $\beta$, at fixed $\Delta=0.7$.  
  Here $\omega=4$ and $\lambda_{D}=0.25$ ($g=2$).}
\label{fig:figS2} 
\end{figure}

\vskip0.20in \noindent
{\bf F: Temperature dependence in the anti-adiabatic regime-}
Fig.~\ref{fig:figS2} shows results at $\Delta=0.7$, near the optimal
disorder, where the pairing susceptibility $\chi_{s}$ is
largest in Fig.~\ref{fig:antiadiabat}.  Unlike Fig.~\ref{fig:figS1},
$S(\pi,\pi)$ no longer grows with $N$ at low temperature, as seen in
Fig.~\ref{fig:figS2}(a).  However, as shown in
Fig.~\ref{fig:figS2}(b), $\chi_{s}$ grows with lattice size,
indicating the presence of robust superconducting correlations in an
intermediate disorder window.  The result of the scaling analysis of
these data is presented in the inset of Fig.~\ref{fig:antiadiabat}.  

The combination of the destruction of CDW order and the rise in SC order illustrated in
the temperature evolution of
Fig.~\ref{fig:figS2}, together with the suppression of $S(\pi,\pi)$
and the onset of $\chi_{s}$ of Fig.~\ref{fig:antiadiabat} 
indicates a competition between the two types of order~\cite{Scalettar89}.
The possibility of a cooperation, in which CDW fluctuations
mediate pairing, has been discussed in~\cite{bickers87}.

\noindent
\begin{figure}[b]
\includegraphics[scale=0.3]{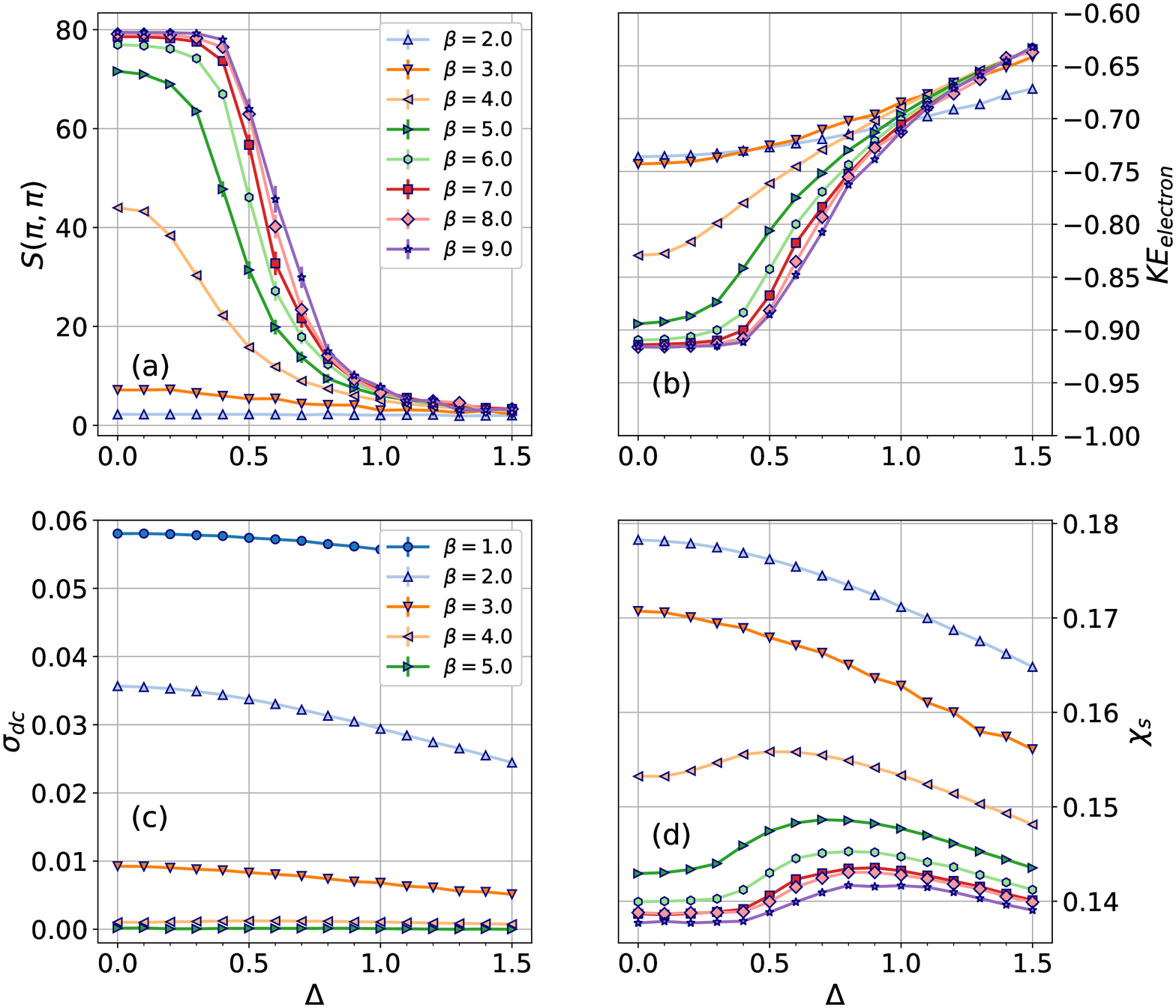}
\caption{Disorder dependence of charge structure factor 
 $S(\pi,\pi)$, the electron kinetic energy $\rm KE_{electron}$, 
 dc conductivity $\sigma_{dc}$ 
 and s-wave pairing susceptibility, panels a-d, respectively 
 at fixed $L=10$.  
 Here $\omega_0=0.5$ and $\lambda_{D}=0.5$ ($g=1$).}
\label{fig:figS3} 
\end{figure}

\vskip0.10in \noindent
{\bf G: Disorder dependence in the adiabatic regime-}
In Fig.~\ref{fig:figS3} we re-plot the data in the adiabatic regime from
Fig.~\ref{fig:adiabat} emphasizing the evolution with disorder.  The
sharp drop in $S(\pi,\pi)$ at $\Delta\sim 0.5$ corresponds to the
destruction of CDW order, with no SC phase.
A further signal of the
transition is seen in the kinetic energy, which becomes smaller in
magnitude upon exiting the CDW phase since virtual hopping is reduced
when sites with electron pairs are no longer surrounded exclusively by
empty sites.

\vskip0.20in \noindent
{\bf H: Analytic Continuation Method-}
We perform the calculation of $A(\omega)$ using the maximum entropy approach
\cite{silver90,gubernatis91,sandvik98}.  This method determines the spectral function
by a weighting which combines a Gaussian piece measuring
the deviation of a computed $G(\tau)$ from the QMC values
for a given $A(\omega)$, and an entropic piece,
with a relative coefficient determined by Bayesian logic.
We use the most straightforward implementation with 
a flat default model (the $A(\omega)$ which would result
in the absence of data), and only the diagonal elements of the covariance
matrix associated with measuring $G$ at two different imaginary time values.

\begin{figure*}[t]
    \includegraphics[width=2\columnwidth]{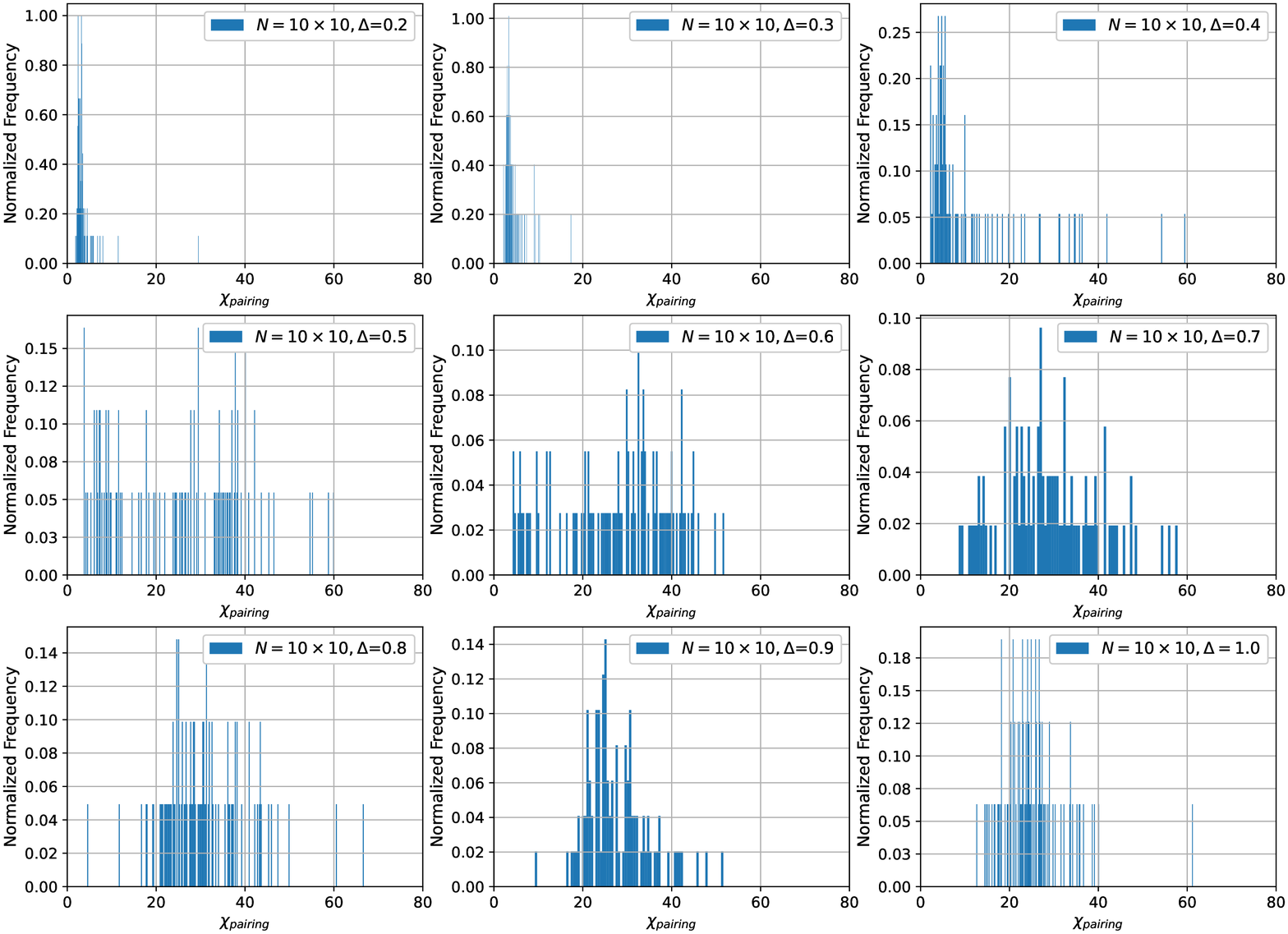}
    \caption{Histograms of distinct realizations of the 
      pairing susceptibility for different disorder strengths $\Delta$.  
      For small $\Delta$, a single narrow peak occurs at small $\chi_{s}$.  
      As $\Delta$ increases, the distribution broadens and shifts to large values.  
      This is the intermediate superconducting phase.  At the largest $\Delta$, 
      the distribution begins returning to smaller values of 
      pairing;  superconductivity is suppressed.
    }
  \label{fig:chihist}
\end{figure*}

\vskip0.10in \noindent
{\bf I: Susceptibility Histograms-}
More detail concerning the enhancement of 
pairing by disorder is given by the 
histograms of the susceptibility
of Fig.~\ref{fig:chihist}.
The figure also gives a sense for the 
realization-to-realization fluctuations
in $\chi_{s}$.

\noindent

\end{document}